\documentclass[a4paper,11pt]{article}
\pdfoutput=1 

\usepackage{jheppub} 
\usepackage{tikz}
\usepackage{dsfont}

\usepackage{hyperref}
\usepackage{cleveref}
\usepackage{supertabular}
\usepackage{todonotes}
\usepackage{mathrsfs}
\usepackage{array}
\usepackage{lipsum}
\usepackage{breqn}
\usepackage{physics}
\usepackage{tikz-cd}
\usepackage{slashed}
\usetikzlibrary{decorations.markings}

\usepackage{helvet} 
\usepackage{amsmath}
\usepackage{amssymb} 

\newcommand{\newreptheorem}[2]{%
\newenvironment{rep#1}[1]{%
 \def\rep@title{#2 \ref{##1}}%
 \begin{rep@theorem}}%
 {\end{rep@theorem}}}
\makeatother

\newreptheorem{lemma}{Lemma}

\newreptheorem{conj}{Conjecture}

\usepackage[most]{tcolorbox} 

\tcbuselibrary{theorems, breakable} 

\newtcbtheorem[number within=section]{definition}{Definition}%
{enhanced,
 breakable, colback=blue!5!white, colframe=blue!75!black,
  fonttitle=\bfseries}{def}

\newtcbtheorem[number within=section]{theorem}{Theorem}%
{enhanced,
 breakable,
 colback=orange!5!white,    
 colframe=orange!85!black,  
 fonttitle=\bfseries,       
 coltitle=white,            
 boxrule=0.8pt,             
 arc=2mm                    
}{thm}

\newtcbtheorem[number within=section]{example}{Example}%
{enhanced,
 breakable, 
 colback=purple!5!white,    
 colframe=purple!100!black,  
 fonttitle=\bfseries,       
 coltitle=white,            
 boxrule=0.8pt,             
 arc=2mm                    
}{ex}

\newtcbtheorem[number within=section]{proposition}{Proposition}%
{enhanced,
 breakable,
 colback=cyan!5!white,    
 colframe=cyan!50!black,  
 fonttitle=\bfseries,       
 coltitle=white,            
 boxrule=0.8pt,             
 arc=2mm                    
}{prop}

\usepackage{lscape} 
\usepackage{braket}

\usepackage[T1]{fontenc} 

\newcommand{\Z}{{\mathbb{Z}}}

\newcommand{\be}{\begin{equation}}
\newcommand{\ee}{\end{equation}}
\newcommand{\ba}{\begin{aligned}}
\newcommand{\ea}{\end{aligned}}
\newcommand{\bea}{\begin{eqnarray}}
\newcommand{\eea}{\end{eqnarray}}

\def\QC{\textrm{\textbf{QC}}}
\def\QCoh{\textrm{\textbf{QCoh}}}
\def\pt{\textrm{pt}}

\def\mb{\mathbb}

\def\mc{\mathcal}

\def\bp{\begin{pmatrix}}
\def\ep{\end{pmatrix}}

\def\bbC{\mathbb{C}}

\def\bbR{\mathbb{R}}

\def\bbZ{\mathbb{Z}}

\def\CG{\mathcal{G}}

\def\mfr{\mathfrak}
\def\Vec{\textrm{\textbf{Vec}}}

\def\mfr{\mathfrak}

\def\Sky{ \mathbf{Sky} }
\newcommand{\ve}[1][]{ #1 \mathbf{Vec}}
\newcommand{\Rep}[1][]{ #1 \mathbf{Rep}}

\title{\boldmath Categorical Continuous Symmetry}
\preprint{}

\author[a]{Qiang Jia}
\author[b,c]{Ran Luo}
\author[d]{Jiahua Tian}
\author[b,e]{Yi-Nan Wang}
\author[f]{Yi Zhang}
\affiliation[a]{Department of Physics, Korea Advanced Institute of Science Technology, \\
Daejeon 34141, Korea}
\affiliation[b]{School of Physics, Peking University,\\
Beijing, China, 100871}
\affiliation[c]{Perimeter Institute for Theoretical Physics,\\
Waterloo, Ontario, Canada N2L 2Y5}
\affiliation[d]{School of Physics and Electronic Science, East China Normal University, \\
Shanghai, China, 200241}
\affiliation[e]{Center for High Energy Physics, Peking University, \\
Beijing, China, 100871}
\affiliation[f]{Kavli IPMU (WPI), UTIAS, The University of Tokyo, \\
Kashiwa, Chiba 277-8583, Japan}
\emailAdd{qjia1993@kaist.ac.kr, ranluo@pku.edu.cn, jtian1905@gmail.com, ynwang@pku.edu.cn, yi.zhang@ipmu.jp}

\abstract{We define the symmetry category in 1+1d for continuous 0-form $G$-symmetry to be $\Sky^\tau(G)$, the category of skyscraper sheaves of finite dimensional vector spaces with finite support on the group manifold of $G$, where $\tau \in H^4(BG,\mathbb{Z})$ is the anomaly. We propose that the corresponding 2+1d SymTFT is described by the Drinfeld center of $\Sky^\tau(G)$. We show explicitly the way that $\tau$ twists the convolution tensor product of the objects of $\Sky^\tau(G)$. As a concrete example, we present the $S$ and $T$-matrices for the simple anyons of the resulting $Z(\Sky^\tau(G))$ category for $G = U(1)$, both for the cases without or with anomaly and discuss the topological boundary conditions as Lagrangian algebra of $Z(\Sky^{\tau}(U(1)))$. We also present the definition of $\Sky^\tau(G)$ and $Z(\Sky^\tau(G))$ for the non-abelian case of $G=SU(2)$, as well as the speculated modular data. We point out that in order to have a physically relevant center and Lagrangian algebras it is necessary to generalize $\Sky^\tau(G)$ to a larger category, which we argue to be closely related to the category of quasi-coherent sheaves on $G_\mathbb{C}$ with convolution tensor product twisted by $\tau$.}

\begin{document}

\maketitle

\section{Introduction}
\label{sec:intro}

There have been a lot of recent developments in the paradigm of Symmetry Topological Field Theory (SymTFT)~\cite{Witten:1998wy,Kong:2020cie,Gaiotto:2020iye,Apruzzi:2021nmk,Hubner:2022kxr,Apruzzi:2022dlm,DelZotto:2022ras,Moradi:2022lqp,Freed:2022qnc,Kaidi:2022cpf,vanBeest:2022fss,Kaidi:2023maf,Chen:2023qnv,Bhardwaj:2023ayw,Apruzzi:2023uma,Cao:2023rrb,Baume:2023kkf,Argurio:2024oym,Copetti:2024onh,Bhardwaj:2024igy,Choi:2024tri,Tian:2024dgl,Najjar:2024vmm,Bonetti:2024etn,Cvetic:2025kdn,Luo:2025phx,Schafer-Nameki:2025fiy,Qi:2025tal,Huang:2023pyk}. In the field theory language, the SymTFT of a $d$-dimensional physical system defined on the $d$-dimensional spacetime, $M_d$, is a TQFT living in a $(d+1)$-dimensional bulk $M_d\times [0,1]$ with two boundaries: (1) the topological boundary, whose boundary condition determines the global structure of the $d$-dimensional physical system, characterized by generalized global symmetries~\cite{Gaiotto:2014kfa,Sharpe:2015mja,Bhardwaj:Lecture,Brennan:2023mmt,Luo:2023ive,Schafer-Nameki:2023jdn}; (2) the dynamical boundary which encodes the local dynamics of the physical system. 

For finite invertible or non-invertible symmetries, there is also a categorical description of the SymTFT~\cite{Kong:2015flk,Ji:2019jhk,ji2022unifiedview,Bhardwaj:2023fca,Bhardwaj:2023idu,Bhardwaj:2023bbf,Hai:2023osv,Bhardwaj:2024qrf,Putrov:2024uor,Chen:2024ulc,Bhardwaj:2024wlr}. We first focus on the case of $d=2$, one first define a ``symmetry category'' $\mc{C}$ that is a fusion 1-category with finitely many simple objects. For instance for a finite symmetry group $G$ with anomaly $\omega\in H^3(G,U(1))$, the symmetry category is given by $\mc{C}=\mathbf{Vec}_G^\omega$, the category of $G$-graded vector spaces with $\omega$ appearing in the definition of its associator.

Then, the SymTFT is described by the Drinfeld center $Z(\mc{C})$, a modular tensor category with finitely many simple objects from the definition. Different topological boundary conditions of the SymTFT would correspond to different Lagrangian algebras in $Z(\mc{C})$, which can be explicitly computed by the modular $S$ and $T$ matrices, e.g. for $\mc{C}=\mathbf{Vec}_G^\omega$. 

On the contrary, for continuous symmetries which appear commonly in nature, the situation is much more complicated. Recently there have been approaches to formulate the SymTFT of continuous symmetry groups as a $BF$-theory with non-compact gauge field $B$ or as a Yang-Mills theory~\cite{Brennan:2024fgj,Antinucci:2024zjp,Bonetti:2024cjk,Apruzzi:2024htg,Antinucci:2024bcm,Gagliano:2024off,Jia:2025jmn,Bonetti:2025dvm}. In particular, the symmetry generator for non-abelian Lie group $G$ in the $(d+1)$-dimensional SymTFT was constructed in \cite{Jia:2025jmn} and there have been recent applications to continuous spacetime symmetries in \cite{Apruzzi:2025hvs,Borsten:2025pvq}. It is natural to wonder if the categorical formulation of SymTFT can be applied to continuous symmetries as well. The fundamental obstacle is the requirements of having finitely many simple objects, in the definition of the fusion category $\mc{C}$ and the modular tensor category $Z(\mc{C})$. To generalize the categorical structure for $\mc{C}=\mathbf{Vec}_G^\omega$ to the case of a Lie group $G$, we need to go beyond the notions of fusion categories, and allowing uncountably infinitely many objects in the category.

For $G=U(1)$, it is imaginable that many results should be consistent with the $\mb{Z}_N$ approximation when $N\rightarrow \infty$. But for non-abelian (semi-simple) Lie groups such as $G=SU(2)$, there is no valid finite group approximation and hence we need brand new techniques to study its categorical aspects.

\paragraph{Main results} 

The main contribution of this work is that we propose that the symmetry category of a 2d bosonic QFT $\mathcal{T}_G^\tau$ with 0-form $G$-symmetry for compact connected Lie group $G$ and 't Hooft anomaly $\tau\in H^4(BG,\mathbb{Z})$ is $\Sky^{\tau}(G)$~\cite{Freed:2009qp}, whose objects are skyscraper sheaves of finite dimensional vector spaces supported on finite subsets of $G$. The monoidal tensor structure of $\Sky^\tau(G)$ is given by a convolution tensor product twisted by $\tau$. For the case of $G=U(1)$, we explicitly derive the twisted convolution with a pair of hermitian line bundle and isometry $(K\rightarrow U(1)\times U(1),\theta)$. More generally for non-abelian $G$, for $X,Y\in \text{Obj}(\Sky^{\tau}(G))$, we have:
\begin{equation}
    X * Y = m_*(\mathscr{M} \otimes p_1^*X \otimes p_2^*Y)
\end{equation}
for a \emph{gerbe bimodule} $\mathscr{M}$ uniquely determined by $\tau$, the precise meanings of which and of other symbols in the above equation will be presented in Section~\ref{sec:General_Remarks_Lie}. 

We explicitly construct the Drinfeld center $Z(\Sky^{\tau}(G))$ in particular for continuous abelian $G$, which is equivalent to the SymTFT corresponding to $\mathcal{T}_G^\tau$, from both the lattice methods in \cite{Freed:2009qp} as $Z(\Sky^{\tau}(G))=\Sky(C)$ for a space $C$, and a group theoretic approach analogous to the cases of finite groups. 

In absence of anomaly $\tau=0$, the simple objects in $Z(\Sky^{\tau}(U(1)))$ are labeled by a pair $(x,z)$ where $x\in U(1)$ is a group element and $z\in\mb{Z}$ labels a irreducible representation of $U(1)$. While for non-zero anomaly of $\tau$ at $k\neq 0$, the simple objects in $Z(\Sky^{\tau}(U(1)))$ are labeled by $(x,z)$ where $x\in U(1)$, $z\in\mb{Z}_{2k}$.

We further derive the braiding structure of $Z(\Sky^{\tau}(G))$ and compute the modular data of $\Sky^\tau(G)$ for $G = U(1)$ with arbitrary anomaly $\tau$, which are consistent with results from the finite group approach.

For the non-abelian case of $G=SU(2)$, we present the simple objects in $Z(\Sky^{\tau}(SU(2)))$ for the case of $\tau=0$ as a pair $(r,n)$ where $r\in[0,2\pi]$ labels the conjugacy classes of $SU(2)$, and $n$ labels the irreducible representation of the centralizer group $C_{SU(2)}(g)$ for the conjugacy class $r$ containing $g$. In absence of anomaly, we also give speculated form of modular matrices for $G = SU(2)$. We then present discussions of Lagrangian algebras for $Z(\Sky^{\tau}(G))$ in these cases, with their physical interpretations, including the gauging of non-normal $\mb{Z}_N$ subgroups for $SU(2)$.

Strictly speaking, the Lagrangian algebra for $Z(\Sky^{\tau}(G))$ should be an object in $Z(\Sky^{\tau}(G))$, which is a sheaf with finite support. Such a condition is not satisfied for Lagrangian algebras with physical relevance. We propose that such subtlety can be resolved by enlarging the category of skyscraper sheaves $\Sky^\tau(G)$ to a category of quasi-coherent sheaves of finite dimensional vector spaces $\mathbf{QC}^\tau(G)$, but we would not finish the full analysis in this paper.

\paragraph{Structure of the paper} 

We present the construction of $Z(\Sky^{\tau}(U(1)))$ in Section~\ref{sec:symcatu1}. We first reformulate the fusion category $\mathbf{Vec}_G^\omega$ for finite $G$ in terms of the ``twisted group ring''~\cite{Freed:2009qp} Vec$_G^\tau$ $(\tau\in H^4(BG,\mb{Z}))$ in Section~\ref{sec:Vec^Tau_G}, to prepare for the generalization to continuous groups later. In Section~\ref{sec:Sky^tau_U1} we review the construction of $\Sky^{\tau}(U(1))$ in \cite{Freed:2009qp} with an explicit computation of the twisted convolution. In Section~\ref{sec:Z(Sky(G))} we construct the Drinfeld center $Z(\Sky^{\tau}(U(1)))$.

In Section~\ref{sec:modulardata} we compute the modular data for $Z(\Sky^{\tau}(U(1)))$ using the limit of $\mb{Z}_N$ $(N\rightarrow\infty)$. We also present the discussion of Lagrangian algebras and match them with the known facts about gauging $\mb{Z}_q$ subgroups in an anomalous $U(1)$.

We discuss the case of non-abelian $G=SU(2)$ in Section~\ref{sec:non-abelian}. In Section~\ref{sec:General_Remarks_Lie} we first give general remarks on constructing the $\Sky^{\tau}(G)$ with multiplicative bundle gerbe. In Section~\ref{sec:SU2} we give a physical derivation of modular data for $G=SU(2)$ and then apply it to a number of Lagrangian algebras. 

Then in Section~\ref{sec:Finiteness} we comment on the technical issue that the objects in $Z(\Sky^{\tau}(G))$ only have finite support, hindering the rigorousness of the definition for the Lagrangian algebras. We propose to solve the issue by introducing the category of quasi-coherent sheaves, but such mathematical object only applies to affine algebraic groups in the current formulations, and it should be generalized in order to apply to the cases of $G=U(1)$ and $SU(2)$. 

In Section~\ref{sec:discussions}, we conclude the main contributions of this paper, and provide discussions on possible future directions regarding the concrete constructions for extended TQFT for the continuous SymTFT, mathematical generalization to justify non-finitely supported objects and extension to fermionic systems.

\section{Symmetry Category of $U(1)$ in $(1+1)$D}
\label{sec:symcatu1}

In this section we propose that the symmetry category of a QFT $\mathcal{T}^\tau_{U(1)}$ with 0-form $U(1)$ symmetry and 't Hooft anomaly characterized by $\tau \in H^4(BU(1),\mathbb{Z})$ is $\Sky^\tau(U(1))$, of which the mathematical description is established in~\cite{Freed:2009qp}. We first review a reformulation of the symmetry category of finite group symmetry from the well-known $\Vec^\omega(G)$ for $\omega \in H^3(G, U(1))$ to the equivalent but relatively less-well-known $\textrm{Vec}^\tau_G$ for $\tau\in H^4(BG, \mathbb{Z})$ in Section~\ref{sec:Vec^Tau_G}. Thanks to the clear geometrical meaning of the monoidal tensor structure of $\textrm{Vec}^\tau_G$, this reformulation paves the way to generalizing the categorical setup to $G = U(1)$, which results in $\Sky^\tau(U(1))$ being the symmetry category of $\mathcal{T}^\tau_{U(1)}$ which will be detailed in Section~\ref{sec:Sky^tau_U1}. In Section~\ref{sec:Z(Sky(G))} we construct the Drinfeld center of $\Sky^\tau(U(1))$, which physically is the SymTFT associated to $\mathcal{T}^\tau_{U(1)}$. In Section~\ref{sec:U1^n} we generalize the construction to $G = U(1)^n$.

\subsection{A Reformulation of the Symmetry Category of Finite Group Symmetry}\label{sec:Vec^Tau_G}

In previous literature, the mathematical description for finite semi-simple categorical symmetries is a fusion $n$-category. Specifically, for finite group symmetries (with possible anomalies) in $(1+1)$-dimensional QFT, the categorical description is given by the category of $G$-graded vector spaces $\ve[ ]_G^\omega$, where $\omega\in H^3(G,U(1))$ signifies the anomaly given by non-trivial associators.

In~\cite{Freed:2009qp}, an alternative and more geometrical (but equivalent) definition of $\ve[]^\omega_G$ in the language of vector bundles was introduced. Following their notation, we denote this category by ${\rm Vec}_G^\tau$, where now $\tau \in H^4(BG,\bbZ)$  as $H^4(BG,\bbZ) \cong H^3(G,U(1))$ for finite group $G$.~\footnote{%
For a finite group $G$, we have $H^\bullet(BG,\mathbb{R})=0$, hence the Bockstein homomorphism $\beta$ from the coefficient short exact sequence $0 \rightarrow \mathbb{Z} \rightarrow \mathbb{R} \rightarrow U(1) \rightarrow 0$ gives an isomorphism $\beta:H^3(G,U(1)) \stackrel{\sim}{\longrightarrow} H^4(BG,\bbZ)$.} This reformulation allows one to extend the scope of categorical descriptions to include Lie groups, specifically $U(1)$, and construct its Drinfeld center. We review the definition of ${\rm Vec}_G^\tau$ and explain the relation to $\ve[]^\omega_G$ when $\beta(\omega) = \tau$.

\begin{definition}{${\rm Vec}_G^\tau$ as a bundle category}{vecgt}\label{def:VecGt}
    The fusion category ${\rm Vec}_G^\tau$ ($G$ is a finite group, $\tau\in H^4(BG,\bbZ)$) is defined with the following data:
    \begin{enumerate}
        \item an object in ${\rm Vec}_G^\tau$ is a complex vector bundle over $G$;
        \item a morphism between two bundles $f\in {\rm Hom}(W,W')$ is a linear vector bundle map;
        \item there is a pair $(K,\theta)$ determined by a given $\tau$ consisting of a hermitian line bundle $K \to G\times G$ and an isometry $\theta_{x,y,z}$ for each $x,y,z\in G$,
        \be\label{eq:thetadef}  \theta_{x,y,z} : K_{y,z}\otimes K^{-1}_{xy,z} \otimes K_{x,yz}\otimes K_{x,y}^{-1}\to \bbC \ee
        satisfying the cocycle condition
        \be \label{eq:thetacocycle} \theta_{y,z,w}\theta^{-1}_{xy,z,w}\theta_{x,yz,w}\theta^{-1}_{x,y,zw}\theta_{x,y,z} = 1 \ , \ \forall x,y,z,w \in G \ .  \ee
        The tensor product $*$ is given by
        \be\label{eq:VecG_convolution} ( W * W')_g =  \bigoplus_{hk = g} K_{h,k} \otimes W_h\otimes W'_k \ . \ee
    \end{enumerate}
\end{definition}
The crucial part of this definition is to convert $H^4(BG,\bbZ)$ into the set of pairs $(K, \theta)$, the bijection between these data is proven in Proposition 4.4 of~\cite{Freed:2009qp}. Concretely, one can take a 3-cocycle $\omega \in H^3(G,U(1))$ (hence also a $\tau$ through $\beta$) which translates to the bundle isometry data $(K,\theta)$ by setting
\begin{equation} \label{eq:33}
    K_{x,y} = \bbC, \quad \text{and} \quad \theta_{x,y,z} = \omega_{x,y,z}\,.
\end{equation}
Conversely, given the pair $(K,\theta)$, we can choose $\{k_{x,y}\in K_{x,y} |x,y\in G\}$ with unit norms and the cohomology class of 
\be\label{eq:34}
\omega_{x,y,z} = \theta_{x,y,z} \big( k_{y,z}k^{-1}_{xy,z} k_{x,yz} k^{-1}_{x,y} \big) \in \bbC^\times  \ , \ \forall x,y,z\in G  \,,
\ee
is independent of the choice of $\{k_{x,y}\}$ and the representatives $(K,\theta)$~\footnote{$(K,\theta)$ can also be viewed as 2-cocycles on $G$ with values in hermitian lines.}. It is then clear that $K$ gives the fusion information and $\theta$ gives the $F$-move information. From this point of view, the equivalence between ${\rm Vec}_G^{\tau=\beta(\omega) }$ and $\ve[]_G^\omega$ is manifest.

\subsection{$\Sky^{\tau}(U(1))$ as the Symmetry Category of $U(1)$ Symmetry}
\label{sec:Sky^tau_U1}

The idea of attaching finite dimensional $\bbC$-vector spaces as fibers to a point of a finite $G$ can be generalized to continuous $G$ by the merit of the sheaf language. With the previous construction, one can now properly generalize the construction of symmetry category into compact abelian Lie groups. In our main reference~\cite{Freed:2009qp}, this is done for a general $n$-Torus $T^n$, we will focus on $G=U(1)$ for the moment.
\begin{definition}{Symmetry category of $U(1)$}{symcatu1} \label{def:skyu1}
    The symmetry category of $U(1)$-symmetric bosonic\footnote{%
    See Appendix~\ref{sec:appA} for an explanation on the level of $U(1)$ anomalies and its relation to spin structures.} theories in $(1+1)$-dimension is a tensor category $\Sky^{\tau}(U(1))$ with $\tau \in H^4(BU(1),\bbZ)$ consisting of the following data:
    \begin{enumerate}
        \item an object in $\Sky^{\tau}(U(1))$ is a skyscraper sheaf of a finite-dimensional vector space on $U(1)$ with finite support. More precisely, picking a finite subset $\{x_1,x_2,\ldots,x_n\}\in U(1)$ and specifying the stalks by finite dimensional vector spaces $\{V_1,V_2\ldots,V_n\}$ at these points define such a sheaf. We will denote simply by $\bbC_x$ the objects with the subscript emphasizing that the skyscraper sheaf is determined by their stalks;
        \item a morphism is a sheaf morphism preserving the linear structure;
        \item a pair of hermitian line bundle and isometry $(K\to U(1)\times U(1),\theta)$, giving the tensor product
        \be\label{eq:SkyU1_convolution} \bbC_x * \bbC_y = K_{x,y} \otimes \bbC_{xy} \ ,  \ee
        and the twist defined as \eqref{eq:thetadef} and satisfying \eqref{eq:thetacocycle}.
    \end{enumerate}
\end{definition}
Here we adopt the notion of a skyscraper sheaf (instead of following the vector space notion as before) with a direct intent of incorporating the information of topology into our discussion. In the cases of finite group, the default setting is to take the discrete topology, however in the Lie group cases, we can choose either discrete topology or continuous topology. In the scope of this paper, we choose continuous topology for Lie groups unless otherwise specified.

In the finite group construction we have explained in \eqref{eq:33} and \eqref{eq:34} how $\tau = \beta(\omega)$ is translated into $(K,\theta)$ (and vice versa), and we intend to explain how one can build $(K,\theta)$ from $\tau \in H^4(BU(1),\bbZ)$ here. We define the lattices of characters and cocharacters
\be\ba
\Lambda &:= {\rm Hom}\big(G,U(1)\big) \cong H^2\big(BU(1),\bbZ\big) \cong \bbZ \subset \mfr{u}(1)^* \\
\Pi &:= {\rm Hom}\big(U(1),G\big) \cong H_2\big(BU(1),\bbZ\big) \cong \bbZ\subset \mfr{u}(1) \ .
\ea\ee
The structure of the cohomology ring $H^\bullet(BU(1),\bbZ)$ gives the following sequence of isomorphisms
\be\label{eq:tauform}\ba
\tau \in H^4(BU(1),\bbZ) &\cong {\rm Sym}^2(\Lambda)\\ &\cong \{ {\rm symmetric \ biadditive \ even \ forms} \expval{\cdot,\cdot}:\Pi\times \Pi\to \bbZ \}\\
&\cong \{ {\rm even \ homomorphisms\ } \tau: \Pi\to \Lambda  \}\,, 
\ea\ee
and from the last line, $\tau$ as an even homomorphism has the explicit formula~\cite{Freed:2009qp} 
\begin{equation}
     \tau (n \cdot  \kappa)   =  2kn \cdot c_1\,, 
\end{equation}
here $\kappa$ and $c_1$ are generators of $\Pi$ and $\Lambda$. $m,n$ and $k$ are integers, where $k$ is also called the level of $\tau$. A detailed discussion about the above isomorphisms and computations can be found in Appendix~\ref{sec:appB}.

The next step is to demonstrate an explicit translation from $\tau\in H^4(BU(1),\bbZ)$ to the $(K,\theta)$ data in implementing the fusion and twist. 

\paragraph{The twisted convolution.}

The strategy is to use $\tau$ to define an action of $\Pi\times \Pi$ on the trivial line bundle on $\mfr{u}(1)\times \mfr{u}(1)$ (universal covering of $\Pi \times \Pi$), and modding out the $\Pi\times \Pi$ action compactifies the trivial bundle on $\mfr{u}(1)\times \mfr{u}(1)$ into a line bundle on $U(1)\times U(1)$, which is the hermitian line bundle $K$ we strive for, and there would be a collection of natural isomorphisms, which is exactly $\theta$. Following the above analysis, a given $\tau\in H^4(BU(1),\bbZ)$ defines an even bilinear form $\expval{\cdot,\cdot}:\Pi\times \Pi\to \bbZ$, which can then be lifted to $\expval{\cdot,\cdot}_{\otimes \bbR}:\mfr{u}(1)\times\mfr{u}(1)\to \bbR$ by tensoring with $\bbR$. We then choose a not necessarily symmetric bilinear form $B:\Pi\times \Pi\to \bbZ$ s.t.
\be  B(\kappa_1,\kappa_2) + B(\kappa_2,\kappa_1) = \expval{\kappa_1,\kappa_2} \ , \ \forall \kappa_{1,2}\in \Pi \ , \ee
and correspondingly we can have $B_{\otimes \bbR}:\mfr{u}(1)\times\mfr{u}(1)\to \bbR$ which can be regarded as a 2-cochain. 
There is a natural action of $\Pi\times \Pi \, \Big(\subset \mfr{u}(1)\times \mfr{u}(1)\Big)$ on $\mfr{u}(1)\times \mfr{u}(1)$ by the simple addition
\be\label{eq:simpleaction} (\kappa_1,\kappa_2) : (\xi_1,\xi_2) \mapsto (\xi_1+\kappa_1,\xi_2+\kappa_2) \ .  \ee
Consider the trivial line bundle $\mfr{K}\to \mfr{u}(1)\times \mfr{u}(1)$, we can lift the previous action of $\Pi\times \Pi$ onto $\mfr{K}$, where in the base space the action is as \eqref{eq:simpleaction} and on the fiber the action is induced by $B_{\otimes\bbR}$
\be\ba 
(\kappa_1,\kappa_2):\mfr{K}_{\xi_1,\xi_2} &\to \mfr{K}_{\xi_1+\kappa_1,\xi_2+\kappa_2}  \\ 
  a  &\mapsto  a\cdot  \exp \Big\{i\pi \big( B_{\otimes\bbR}(\kappa_1,\xi_2)-B_{\otimes\bbR}(\kappa_2,\xi_1) + B_{\otimes\bbR}(\pi_1,\pi_2) \big) \Big\} \ .
\ea\ee
Taking quotient of the $\Pi\times \Pi$ action, the base space becomes $U(1)\times U(1)$, and the fiber is still $\bbC$. The quotient is the hermitian line bundle $K\to U(1)\times U(1)$ we desire. Although the construction depends on the choice of $B$, Corollary 2.9 in~\cite{Ganter:2014zoa} shows that the equivalence class of $\Sky^{\tau}(U(1))$ is in fact independent of $B$.

Now on $U(1)^{\times 3}$, for every $x,y,z\in U(1)$, there are two different ways to tensor the line bundles pulled back from $K$, namely
\be  K_{y,z}\otimes K_{x,yz}  \ {\rm and} \ K_{x,y}\otimes K_{xy,z} \ , \ee
and they must be isomorphic. The isomorphism is denoted \be\theta_{x,y,z}: K_{y,z}\otimes K_{x,yz} \to K_{x,y}\otimes K_{xy,z} \ ,  \ee
or equivalently
\be  \theta_{x,y,z}: K_{y,z}\otimes K^{-1}_{xy,z}\otimes K_{x,yz}\otimes K^{-1}_{x,y} \to  \bbC \ .  \ee

\subsection{Construction of the Drinfeld Center of $\Sky^\tau(U(1))$}\label{sec:Z(Sky(G))}

An object in the Drinfeld center $Z\Big(\Sky^{\tau}(U(1))\Big)$ shall be in the form of $ ( V , \gamma_V  ) $ where $V \in {\rm Obj}\Big(\Sky^{\tau}(U(1))\Big)$ and $\gamma_V$ being a natural isomorphism, for each $W\in {\rm Obj}\Big(\Sky^{\tau}(U(1))\Big)$ it gives an isomorphism
\be \gamma_{V,W} : V\otimes W \stackrel{\sim}{\longrightarrow} W\otimes V \ . \ee
Down to the level of simple objects, $\bbC_x \ \big(x\in U(1)\big)$, we observe that
\be\ba
&\bbC_x * \bbC_y = K_{x,y} \otimes \bbC_{xy}\\
&\bbC_y * \bbC_x = K_{y,x} \otimes \bbC_{xy}\\
\Rightarrow \quad & K_{x,y}^{-1}\otimes K_{y,x} \big( \bbC_x * \bbC_y \big) =  \bbC_y * \bbC_x \ ,
\ea\ee
in other words, we have the isomorphism
\be  K_{x,y}^{-1}\otimes K_{y,x} :  \bbC_x * \bbC_y \stackrel{\sim}{\longrightarrow}  \bbC_y * \bbC_x \ . \ee
To give a more concise description of $Z\Big(\Sky^{\tau}(U(1))\Big)$, we can picture a space $C$ with $\Sky\big(C\big)$ being the Drinfeld center, it should be intuitively constructed as
\be
\begin{tikzcd}
     & C \arrow[ld, "p_1"'] \arrow[rd, "p_2"] &                          \\
U(1) &                                        &  X
\end{tikzcd}
\ee
where $p_{1,2}$ are projections, and the projection to some space $X$ shall give the $K_{x,-}^{-1}\otimes K_{-,x}$ information for each $x\in U(1)$. Following \cite{Freed:2009qp}, the space $C$ is constructed as
\be\ba
C &= \big(\mfr{u}(1) \oplus \Lambda  \big)/\Pi \\
X &= \Lambda/\tau(\Pi)
\ea\ee
 where the action of $\kappa\in\Pi$ on $(\xi , \lambda) \in \big(\mfr{u}(1) \oplus \Lambda  \big)$ is induced by $\tau$, namely
\be  \kappa \cdot (\xi , \lambda) = (\xi + \kappa, \lambda + \tau(\kappa))  \ . \ee
where we have taken $\tau$ to be an even homomorphism $\tau:\Pi\to \Lambda$, as explained in \eqref{eq:tauform}.

\begin{proposition}{Drinfeld center of $\Sky^{\tau}(U(1))$~\cite{Freed:2009qp} }{}
    The category $\Sky\big(C\big)$ constructed above is the Drinfeld center of $\Sky^{\tau}(U(1))$.
\end{proposition}

Now we specify the simple objects of the Drinfeld center $Z\big(\Sky^{\tau}(U(1))\big)\cong \Sky(C)$.

\begin{itemize}
    \item In the non-anomalous case, we have $\tau =0$, thus $C = U(1)\oplus \Lambda$, a simple object in $\Sky(C)$ is labeled by
    \be (x, z)  \ , \ x\in U(1) \ , \ z\in \Lambda\cong {\rm IrRep}(U(1)) \ .  \ee
    Note that the classification of simple objects of $Z(\ve[]_G)$ in the finite group case is given by
    \be (x, \rho) \ , \ x\in {\rm Cl}(G) \ , \ \rho\in {\rm IrRep}(C_x G)  \ee
    where $C_xG$ is the centralizer of $G$ with respect to $x$. By naively taking $G$ to be $U(1)$, the classification of simple objects aligns with the result of $Z\big(\Sky^{\tau}(U(1))\big)\cong \Sky(C)$.
    \item In the anomalous case, we have $\tau$ at level $k \ne 0$, inducing a non-trivial action of $\Pi$ on $\mfr{u}(1)\oplus \Lambda$ given by
    \be \kappa\cdot (\xi,\lambda) =(\xi + \kappa , \lambda + 2k\kappa ) \ , \ \forall \kappa\in \Pi \ , \ \xi \in \mfr{u}(1) \ , \ \lambda\in\Lambda  \ee
    Modding out this action, we obtain the result that a simple object in $\Sky(C)$ is labeled by
    \be (x, z) \ , \ x\in U(1) \ , \ z\in \bbZ_{2k} \ .   \ee
    Note that this result is in alignment with the result obtained from finite group approximation, as we will briefly mention in the following and present the details in Section \ref{sec:modulardata}.
\end{itemize}
The braiding between simple objects $X=[(\xi_1, \lambda_1)]$ and $Y=[(\xi_2, \lambda_2)]$ is computed only in part~\cite{Freed:2009qp}--i.e. only the $\mathfrak{u}(1)$-part--although the authors indicate how to carry out the full computation. It is straightforward to get the full result as the $S$-matrix
\begin{equation}
 S_{X,Y} = R_{X,Y} R_{Y,X} = \exp \Big( 2 \pi i \left(\xi_1 \lambda_2 + \xi_2\lambda_1 - 2 k \,\xi_1 \xi_2 \right) \Big) \,,
\end{equation}
where 
\begin{equation}
R_{X,Y} = \exp \Big( 2\pi i (\lambda_1\xi_2 -  B_{\otimes\bbR}(\xi_2,\xi_1))   \Big)
\end{equation}
is the braiding, compare Proposition 3.5.0.1 of~\cite{Weis:2022egw} for an interpretation via \emph{categorical tori}.

\paragraph{Simple objects and $S/T$-matrices by finite approximation.} 
 We also have a description of the objects in $Z\big(\Sky^{\tau}(U(1))\big)$ analogous to the cases of finite group~\cite{Freed:2009qp, naidu2008lagrangian}. 
 Since $U(1)$ can be approximated by its cyclic subgroup $\bbZ_N$ as $N \rightarrow \infty$, the subtle issue here is that this limit does not yield  $U(1)$ with its standard topology, but rather $U(1)^\delta$, i.e. $U(1)$ equipped with discrete topology. At the level of $U(1)^\delta$, an anomaly $\tau \in H^4(BG,\bbZ)$ corresponds to a class in $H^3(BU(1)^\delta,U(1))$, in much the same way that a Chern--Simons invariant associates to $\tau$. In fact, the \emph{Eilenberg-MacLane} space 
 \begin{equation}
     BU(1)^\delta = K(U(1),1)
 \end{equation}
serves as the classifying space of flat $U(1)$-bundles. What we describe here only scratches the surface; further details on anomaly and flat gauging will be presented in another article~\cite{fiveguys_anomaly}.
The quantities of $Z\big(\Sky^{\tau}(U(1))\big)$ obtained by finite approximation we be presented in Section~\ref{sec:modulardata}.

\subsection{Generalization to $U(1)^n$}
\label{sec:U1^n}

The above construction can be generalized to the case when the symmetry is $U(1)^n$ with simple modifications.

The symmetry category of $U(1)^n$ with anomaly $\tau\in H^4(BG,\bbZ)$ is given by:
\begin{definition}{Symmetry category of $U(1)^n$}{symcatu1n}
    The symmetry category of $U(1)^n$ symmetry in $(1+1)$-dimension is $\Sky^\tau(U(1)^n)$ ($\tau \in H^4(BU(1)^n,\bbZ)$), consisting of the following data:
    \begin{enumerate}
        \item an object in $\Sky^\tau(U(1)^n)$ is a skyscraper sheaf of finite-dimensional vector spaces on $U(1)^n$ with finite support;
        \item a morphism is a sheaf morphism preserving the linear structure;
        \item a pair of hermitian line bundle and isometry $(K\to U(1)^n\times U(1)^n,\theta)$ determined by $\tau$, giving the tensor product
        \be \bbC_x * \bbC_y = K_{x,y} \otimes \bbC_{xy} \ ,  \ee
        and the twist defined as \eqref{eq:thetadef} and satisfying \eqref{eq:thetacocycle}.
    \end{enumerate}
\end{definition}

Similarly, in the construction of the Drinfeld center, we expect the information of
\be \gamma_{V,W} :V\otimes W \stackrel{\sim}{\longrightarrow} W\otimes V  \ee
where $V,W\in {\rm Obj}\Big( \Sky^\tau(U(1)^n)  \Big)$. In this case, giving simple objects $\bbC_x$ and $\bbC_y$, the information of this natural transformation is again given by
\be K^{-1}_{x,y} \otimes K_{y,x}:\bbC_x * \bbC_y \to \bbC_y * \bbC_x \ . \ee
To incorporate the above information in the Drinfeld center $Z(\Sky(U(1)^n))\cong\Sky(C)$ we are about to construct, we define
\be
\ba
\Pi&\equiv {\rm Hom}(U(1),U(1)^n) \cong H_2(BU(1)^n,\bbZ) \\ 
\Lambda&\equiv {\rm Hom}(U(1)^n,U(1)) \cong H^2(BU(1)^n,\bbZ) \ ,
\ea
\ee
and since \eqref{eq:tauform} still stand, we can regard $\tau\in H^4(BU(1)^n,\bbZ)$ as an even homomorphism $\tau:\Pi\to \Lambda$. The space $C$ is given by
\be
\begin{tikzcd}
     & C \arrow[ld, "p_1"'] \arrow[rd, "p_2"] &                          \\
U(1)^n &                                        &  \Lambda/\tau(\Pi)
\end{tikzcd}
\ee
or in other words, $C = ( \mfr{u}(1) \oplus \Lambda )/\Pi$, where the $\Pi$ action is given by
\be  \pi\cdot (\xi,\lambda  ) = (\xi + \pi, \lambda + \tau(\pi)) \ . \ee
It follows from \cite{Freed:2009qp} that:
\begin{proposition}{Drinfeld center of $\Sky^\tau\big(U(1)^n\big)$ }{}
    The category $\Sky\big(C\big)$ constructed above is the Drinfeld center of $\Sky^\tau\big(U(1)^n\big)$.
\end{proposition}

\section{Modular Data of $\Sky^\tau(U(1))$ through finite group approximation} 
\label{sec:modulardata}

The construction of the symmetry category $\Sky^{\tau}(U(1))$ and the corresponding Drinfeld center $\Sky(C)$, the paramount goal is to understand the structure of the Drinfeld center, which gives a detailed description of topological line operators in the $(2+1)$D SymTFT. 

In this section, we unravel the braiding, topological spin and boundary condition information of $Z(\Sky^{\tau}(U(1)))$ through finite group approximation. The process is to start from the $S,T$ matrices of $Z(\ve[]_{\bbZ_N}^\omega)$, and then take the limit $N\to \infty$ to obtain the result of $Z(\Sky^{\tau}(U(1)))$. 
There is a precise agreement between the finite group approximation and the results obtained in Section~\ref{sec:Z(Sky(G))}, where the analysis is done for the continuous $U(1)$. We verify that the $S,T$ matrices we obtain satisfy the constraints required for $Z(\Sky^{\tau}(U(1)))$ to be a ``modular category'', with the finiteness condition relaxed in the definitions of modular tensor category. We also calculate the Lagrangian algebras constituting the boundary conditions of the SymTFT. Additional details of the proofs of the properties discussed in this section are collected in Appendix~\ref{sec:appD}.

\subsection{Modular Matrices}
\label{sec:modular-U(1)}

 Let us come back to the analysis with finite approximation of $U(1)$. The objects are labeled by $(x,m)$ where\footnote{%
 Note that we have a different parametrization of $U(1)$ and normalize its group volume as $|U(1)|= 2\pi$.} $x=e^{i\theta_x}\in U(1)$ $(\theta_x\in[0,2\pi))$ and $m\in\mb{Z}$ labels a projective representation $\pi_m$ of $U(1)$ subject to the condition
\be
\label{pim-cond}
\pi_m(y)\pi_m(z)=\pi_m(yz)\alpha_x(y,z)\,.
\ee
Here $\alpha_x\in Z^2(U(1)^\delta,U(1))$ is related to the group cohomology element $\omega\in H^3(U(1)^\delta,U(1))$ corresponding to the anomaly with coefficient\footnote{%
$\omega$ for $U(1)^\delta$ can be approximated by the $\omega \in H^3(B\bbZ_N,U(1))$ of an embedded $\bbZ_N$ as $N \rightarrow \infty$, see Eq.~\eqref{eq:anomalyrelationk}. We will provide a derivation from quantum field theory in \cite{fiveguys_anomaly}.} $k$ as
\be
\omega(e^{i\theta_x},e^{i\theta_y},e^{i\theta_z})=\exp\left(-ik\theta_x(\theta_y+\theta_z-[\theta_y+\theta_z])/(2\pi)\right)
\ee
as
\be
\alpha_x(y,z)=\frac{\omega(y,z,(yz)^{-1}x(yz))\omega(x,y,z)}{\omega(y,y^{-1}xy,z)}=\omega(x,y,z)\,.
\ee
$[\theta]$ means $\theta\ \text{mod}\ 2\pi$.

The projective representation satisfying (\ref{pim-cond}) is given by
\be
\pi_m(e^{i\theta})=\exp\left(i\left(m-\frac{k\theta}{2\pi}\right)[\theta]\right)\,.
\ee
From this expression, we can also derive the modular $S$ and $T$-matrices from the standard formula~\cite{naidu2008lagrangian}
\be
S_{(x,n),(y,n')}=\frac{1}{|G|}\pi^*_n(y)\pi^*_{n'}(x)=\frac{1}{2\pi}e^{-i(n\theta_y+n'\theta_x-\frac{2k}{2\pi}\theta_x\theta_y)}\,.
\ee
\be
T_{(x,n),(y,n')}=\delta_{x,y}\delta_{n,n'}\pi_n(x)=\delta_{x,y}\delta_{n,n'}\exp\left(i\left(n\theta_x-\frac{k\theta_x^2}{2\pi}\right)\right)\,,
\ee
where $\delta_{x,y}$ is defined as
    \begin{equation}
        \delta_{x,y}= \sum_{n\in \mathbb{Z}} \delta(\theta_x-\theta_y+2\pi n)\,.
    \end{equation}
The factor $1/|G|$ for finite groups $G$ would go to 0 when $G$ is infinite. For $G=U(1)$ we will normalize $|G|=|U(1)|=2\pi$ as its group volume.
From the forms of $S$ and $T$-matrices, we observe that the lines have the equivalent relation $(\theta_x,n)\sim (\theta_x+2\pi,n+2k)$, as the relevant $S$ and $T$-matrix elements are equalized. We also provide another derivation in Appendix~\ref{sec:appC}, by taking the $N\rightarrow \infty$ limit of the $S,T$-matrices of $Z(\ve[]_{\bbZ_N}^\omega)$.

One can check $S$ and $T$ are unitary
    \begin{equation}
        (S^{\dagger} S)_{(x,n),(y,n')}=(T^{\dagger} T)_{(x,n),(y,n')} =   \delta_{x,y}\delta_{n,n'}\,,
    \end{equation}
and satisfy the following properties 
    \begin{equation}
        (S^2)_{(x,n),(y,n')} = (ST)^3_{(x,n),(y,n')} = C_{(x,n),(y,n')}\,,
    \end{equation}
where the charge conjugation matrix is
    \begin{equation}
        C_{(x,n),(y,n')} =  \delta_{x,y^{-1}}\delta_{n',-n+2k\left(\frac{\theta_x+\theta_y}{2\pi}\right)}\,,
    \end{equation}
and the multiplication between two infinite dimensional matrices $A$ and $B$ is defined as
    \begin{equation}
        (AB)_{(x,n),(y,n')} = \int_{0}^{2\pi} d\theta_z \sum_{n''} A_{(x,n),(z,n'')} B_{(z,n''),(y,n')}\,.
    \end{equation}
The fusion coefficient reads
    \begin{equation}
        N^{(z,n'')}_{(x,n),(y,n')}=  \delta_{xy,z}\delta_{n'',n+n'- 2k \left(\frac{\theta_x+\theta_y-\theta_z}{2\pi}\right)}\,,
    \end{equation}
which gives the fusion rule of anyon lines $W_{(x,n)}$
    \begin{equation}
        W_{(x,n)}\times W_{(y,n')} = \int_0^{2\pi} d \theta_z \sum_n N^{(\theta_z,n'')}_{(\theta_x,n),(\theta_y,n')} W_{(z,n'')} = W_{\left(xy,n_1+n_2- \frac{2k}{2\pi}(\theta_x+\theta_y-[\theta_x+\theta_y])\right)}\,,
    \end{equation}
In particular, setting $\theta_y=n'=0$ we can recover the equivalency $(\theta_x,n)\sim (\theta_x+2\pi,n+2k)$, which aligns with the classification of simple objects in $\Sky(C)=Z(\Sky^{\tau}(U(1)))$.

\subsection{Gauging Cyclic Subgroups and the Lagrangian Algebra}
\label{sec:Gauging_Cyclic}

We now discuss the Lagrangian algebra. First of all, to implement Dirichlet boundary condition, we have
    \begin{equation}\label{eq:LDir}
        \mathcal{L}_{\textrm{Dir}} = \bigoplus_n W_{(1,n)}\,,
    \end{equation}
for any $k$. 

\paragraph{Anomaly-free $U(1)$.} If $k=0$, we further have the following Lagrangian algebra implementing Neumann boundary condition:
\begin{equation}\label{eq:LNeu}
    \mathcal{L}_{\textrm{Neu}} = \bigoplus_{x} W_{(x,0)}\,,
\end{equation}
which corresponds to (flat) gauging $U(1)$, and Lagrangian algebras
\begin{equation}\label{eq:Lq}
    \mathcal{L}_{q} =  \bigoplus_{m} \bigoplus_{n=0}^{q-1} W_{\left(e^{\frac{2\pi i}{q}n}, qm\right)}\,,
\end{equation}
which corresponds to gauging the subgroup $\mathbb{Z}_q\in U(1)$ ($q>1$) which is automatically anomaly free.

We note that~(\ref{eq:LDir}),~(\ref{eq:LNeu}) and~(\ref{eq:Lq}) all involve summing over an infinite family of objects of $\Sky(C) = Z(\Sky(U(1)))$ using the natural additive structure of $\Sky(C)$ as an abelian category. This is clearly in tension with the requirement that the objects of $\Sky(U(1))$ are finitely supported (Data 1 of Definition~\ref{def:skyu1}). Actually, there would be no reason to impose this finiteness condition on $\Sky(U(1))$ if we were about to relax it for $\Sky(C)$. Moreover,~(\ref{eq:LNeu}) is technically even more subtle because it virtually involves summing over an uncountably infinite family of objects. Since it is very natural to be able to impose the boundary conditions associated to the choices~(\ref{eq:LDir}),~(\ref{eq:LNeu}) and~(\ref{eq:Lq}), the issue of summing over both countably and uncountably infinite families of objects is unavoidable as long as one is allowed to gauge the corresponding symmetries. Clearly, this finiteness issue cannot be resolved within the realm of $\Sky(C)$, rather one has to generalize $\Sky(C)$ to a larger category $\mathscr{C}(C)$ in which objects like $\mathcal{L}_{\textrm{Dir}}$, $\mathcal{L}_{\textrm{Neu}}$ and $\mathcal{L}_q$ are well-defined. Therefore $\Sky(C)$ should be a subcategory of $\mathscr{C}(C)$ and it will be even more desirable that the monoidal tensor structure of $\mathscr{C}(C)$ descends to that of $\Sky(C)$ by restriction without further modification. This in turn forces us to consider the corresponding $\mathscr{C}(G)$ with Drinfeld center $Z(\mathscr{C}(G)) = \mathscr{C}(C)$. We will stick to $\Sky(G)$ (and $\Sky^\tau(G)$) and ignore this issue for the moment and come back to this point in Section~\ref{sec:Finiteness} where we propose a candidate of $\mathscr{C}$.

\paragraph{Anomalous $U(1)$.} If $k\neq 0$, the entire $U(1)$ is anomalous. We would then like to look for anomaly free cyclic subgroups $\mathbb{Z}_q$, and this happens if and only if $q>0$ is a divisor of $k$. The reason is the following. In our convention, $\tau = k (c_1)^2$ corresponds to $U(1)$ current algebra at level $2k$. In Section 3.3 of~\cite{Okada:2025kie} the anomaly of an embedded $\mathbb{Z}_q \subset U(1)_{2k}$ subgroup is computed using the standard Else--Nayak's argument~\cite{Else:2014vma}. The anomaly $\tau$ on the subgroup appears as a 3-cocycle with an explicit simple expression 
\begin{equation} \label{eq:anomalyrelationk}
    \omega(e^{2\pi i \frac{a}{q}}, e^{2\pi i \frac{b}{q}}, e^{2\pi i \frac{c}{q}}) = \exp \left(2\pi i \frac{k}{q} a \,p(b,c)\right) \,,
\end{equation}
where $a,b,c \in \{0,1,2,\ldots,q-1\}$ and $p(b,c)$ is defined as
\begin{align}
p(b,c)=\left\{\begin{array}{ll}
0 & (b+c < n)  \\
1 & (b+c \geq n)
\end{array}\right.\,.
\end{align}
We see from the above that $\omega$ is trivial only when $\frac{k}{q} \in \bbZ$, hence in these cases we can gauge the anomaly-free $\mathbb{Z}_q$ subgroup and the corresponding Lagrangian algebra is
    \begin{equation}
        \mathcal{L}^k_q = \bigoplus_m \bigoplus_{n=0}^{q-1} W_{\left(e^{\frac{2\pi i}{q}n},\frac{k}{q}n+qm\right)}\,.
    \end{equation}

Moreover, if $k$ is even there also exists fermionic Lagrangian algebras. When $k=2\ \textrm{mod}\ 4$ one has
    \begin{equation}
        \mathcal{L}_{f} = \bigoplus_{m}\bigoplus_{n=0,1} W_{((-1)^n,2m)}\,,
    \end{equation}
and when $k=0\ \textrm{mod}\ 4$ one has
    \begin{equation}
        \mathcal{L}_{f} = \bigoplus_{m}\bigoplus_{n=0,1} W_{((-1)^n,2m+n)}\,.
    \end{equation}
One can check that the topological spin for each component is $0$ or $\frac{1}{2}$, and they also braid trivially with each other.

\section{Symmetry Category of Non-abelian Continuous Group}
\label{sec:non-abelian}

In this section, we generalize the construction of $\Sky^\tau(G)$ from $G = U(1)^n$ to a compact, simply-connected and simple Lie group $G$. The key is to apply the construction of \emph{manifold tensor category} on $G$, which lets us define $\Sky^\tau(G)$ with the desired monoidal tensor structure. Since the definition of $\Sky^\tau(G)$ is almost in total parallel to Definition~\ref{def:symcatu1} for $G = U(1)$, we will not bother repeating it here. Rather, we will simply remark on the classification of $\tau$ by $H^4(BG,\mathbb{Z}) \cong \bbZ$ (or equivalently the line bundle $K$ playing the role of convolution kernel) and on certain physical subtleties regarding the distinction between flat gauging and dynamical gauging in Section~\ref{sec:General_Remarks_Lie}. The readers may consult~\cite{weis2022a} for an exposition to the rather new idea of manifold tensor categories, where discussions on more general Lie group are covered. In Section~\ref{sec:SU2} we will compute certain modular data of $\Sky(G)$ for $G = SU(2)$.

\subsection{General Remarks}
\label{sec:General_Remarks_Lie}

The framework of manifold tensor category recently developed in~\cite{weis2022a} is particularly useful in developing the symmetry category of a large class of symmetry groups $G$, including discrete groups and general Lie groups. The gist is to construct a tensor category equipped with a (twisted) convolution tensor product whose simple objects ``form a manifold'' (cf. Definition~\ref{def:VecGt} and~\ref{def:symcatu1}). The finiteness condition (cf. Data 1 in Definition~\ref{def:symcatu1}) makes the transcription from $\text{Vec}^\tau_G$ to $\Sky^\tau(G)$ for discrete $G$ verbatim, and the generalization to continuous $G$ immediate. 

More precisely, the monoidal tensor category $\Sky^\tau(G)$ for a Lie group $G$ can be defined in total parallel with Definition~\ref{def:symcatu1} for $G = U(1)$~\cite{weis2022a}, once the pair $(K,\theta)$ has been identified. We emphasize that we always restrict ourselves to skyscraper sheaves that are $C^\infty$-modules supported on a finite set of $G$ as this finiteness condition is crucial in defining the monoidal structure of $\Sky^\tau(G)$. We will discuss some potential generalizations to relax this finiteness condition in Section~\ref{sec:Finiteness}. To further define the monoidal tensor structure of $\Sky^\tau(G)$, the key is to generalize the convolution tensor product~(\ref{eq:SkyU1_convolution}) properly. Concretely, the convolution of two objects of $\Sky(G)$ is:
\begin{equation}\label{eq:convolution_product}
    X * Y = m_*(p_1^*(X)\otimes p_2^*(Y))
\end{equation}
for $m,p_1,p_2 : G\times G\rightarrow G$ where $m$ is the multiplication and $p_1$ and $p_2$ are projections~\cite{polishchuk2011kernel, diFiore_thesis}. Given Definition~\ref{def:VecGt}, it is natural to consider the following twisted convolution as a generalization of~(\ref{eq:convolution_product}):
\begin{equation}\label{eq:twisted_convolution_product}
    X * Y = m_*(p_1^*(X)\otimes p_2^*(Y)\otimes K)
\end{equation}
for a line bundle $K \rightarrow G\times G$.~\footnote{It is actually more than a line bundle. More precisely, it is a \emph{gerbe bimodule}, although it can be viewed as a $U(1)$-bundle (and further approximated by a line bundle) given a trivialization of a certain bundle gerbe on the group manifold of $G$. We will cover this issue in a moment.} The convolution~(\ref{eq:twisted_convolution_product}) is exactly the twisted (by $K$) convolution tensor product on $\Sky^\tau(G)$~\cite{weis2022a}. The line bundle $K$ (together with its $\theta$-profile) is clearly by no means an arbitrary element of $\text{Pic}(G\times G)$, rather, as pointed out in~\cite{weis2022a}, it has to be associated to a \emph{multiplicative bundle gerbe} satisfying the pentagon axiom defined in~\cite{waldorf2010multiplicative}. Proposition 2.8 in~\cite{waldorf2010multiplicative} further implies that for compact $G$ each $\tau \in H^4(BG, \mathbb{Z})$ characterizes a multiplicative bundle gerbes (also see Remark 2.4.2.4 of~\cite{weis2022a}), i.e., the isomorphism classes of these bundle gerbes are classified by $H^4(BG, \mathbb{Z})$~\cite{Carey:2004xt, waldorf2010multiplicative}, which matches the classification of the convolution kernel represented by $K$ in~\cite{Freed:2009qp}. Therefore, for theory with ('t Hooft) anomalous $G$-symmetry the corresponding symmetry category is $\Sky^\tau(G)$, where $\tau \in H^4(BG, \mathbb{Z})$ specifies a twisted convolution tensor product in terms of the line bundle $K$ (with a trivialization of the group manifold of $G$) entering~(\ref{eq:twisted_convolution_product}).

To concretely define $\Sky^\tau(G)$ analogous to the $U(1)$ case, we ought to compose an eloquent translation between cohomology information $\tau$ and bundle information $(K,\theta)$. In the $U(1)$ case, we gave an explicit yet specific construction. As mentioned earlier, the bundle information $(K,\theta)$ can be concisely packed in the mathematical concept of multiplicative bundle gerbe, which we will briefly introduce here. The definition of bundle gerbe~\cite{Murray:1994db} is given by the following.\footnote{\label{fn:gerbe}
The set of data needed to define a gerbe (or ``U(1)-gerbe'' and ``abelian gerbe'') that is rather accessible to physicists is listed as follows~\cite{Alvarez:1984es}: 
\begin{itemize}
    \item A manifold $M$ with a good cover $\bigcup \,\mathcal{U}_i$;
    \item A collections of smooth functions $h_{i j k}:\mathcal{U}_{ijk} \longrightarrow U(1)$
for every non-empty triple intersection $\mathcal{U}_{ijk} := \mathcal{U}_i \cap \mathcal{U}_j \cap\mathcal{U}_k$;
    \item For every non-empty quadruple intersection $\mathcal{U}_{ijkl} := \mathcal{U}_i \cap \mathcal{U}_j \cap\mathcal{U}_k \cap \mathcal{U}_l$ the cocycle condition  $h_{jkl} \, h_{ikl}^{-1} \, h_{ijl} \, h_{ijk}^{-1} = 1$ is satisfied;
    \item Two gerbes $\{h_{i j k}\}$ and $\{h'_{i j k}\}$ are equivalent if there exist a collection of smooth functions $\{f_{i j}| f_{i j}: \mathcal{U}_{ij} := \mathcal{U}_i \cap \mathcal{U}_j: \longrightarrow U(1), f_{j i} = f^{-1}_{i j} \;\text{and}\; f_{i i}=1\}$, such that $h'_{i j k} = h_{ijk} \cdot f_{i j} f_{j k} f_{k i}$ on $\mathcal{U}_{ijk}$; 
    \item Isomorphic classes of gerbes is classified by $H^3(M,\bbZ)$ and the corresponding image of $[\{h_{ijk}\}]$ in $H^3(M,\bbZ)$ is called its \emph{Dixmier-Douady} class.
\end{itemize}
 We take the submersion as the disjoint union $Y=  \bigsqcup_i \mathcal{U}_i$. Then we recognize immediately that $Y^{[2]} = \bigsqcup_{i,j} U_{ij}$, $Y^{[3]} = \bigsqcup_{i,j,k} U_{ijk}$, $Y^{[4]} = \bigsqcup_{i,j,k,l} U_{ijkl}$ and etc. We can now take trivial $U(1)$-bundles $L_{i j}$ on each $U_{ij}$, and then glue them to a line bundle $L$ over $Y^{[2]}$ using gerbe cocycle. Next define 
\begin{equation}
\begin{aligned}
        \mu: L_{i j} \otimes L_{j k} &\rightarrow L_{i k} \\
        (x,v) \otimes(x,w)&\mapsto (x,h_{ijk}(x)vw) \,.
\end{aligned}
\end{equation}
This multiplication satisfies the associativity condition automatically because $h_{ijk}$ satisfies the quadruple-overlap cocycle condition. Conversely, given a bundle gerbe on $M$, one can recover the gerbe described above. A key advantage of the geometric model $(Y,L,\mu)$ is that it also applies to manifolds that may not admit good covers. For Lie groups, bundle gerbes and gerbes are therefore equivalent.}
\begin{definition}{Bundle gerbe}{}
    A bundle gerbe $\CG$ over a smooth manifold $M$ is given by the following data:
    \begin{enumerate}
        \item a surjective submersion $\varphi:Y\to M$;
        \item a $U(1)$-principle bundle
        \be L\to Y\times_M Y  \ee
        over the fiber product;\footnote{%
        $Y^{[2]}:=Y \times_M Y = \{(y_1, y_2) \in Y^2 \mid \varphi(y_1)=\varphi(y_2)\}$, higher fiber products are defined similarly by adding $Y$ factors.}
        \item an isomorphism of $U(1)$-bundles on $Y^{[3]} := Y\times_M Y\times_M Y$,
        \be  \mu: L_{12} \otimes L_{23} \to  L_{13} \ee
        where $L_{i j}$ is the pullback of $L$ to the $(i,j)$-th factor of $Y^{[3]}$; 
        \item the isomorphism is associative over $Y^{[4]} := Y\times_M Y\times_M Y\times_M Y$, namely,
        \be \mu_{1,3,4}\circ(\mu_{1,2,3}\otimes {\rm id} )  = \mu_{1,2,4} \circ( {\rm id}\otimes \mu_{2,3,4} ) \ .\ee
    \end{enumerate}
\end{definition}

 With the additional multiplication structure $m:G\times G\to G$, we can define a \textbf{multiplicative bundle gerbe}. 
\begin{definition}{Multiplicative bundle gerbe}{}
A multiplicative bundle gerbe $(\mathcal{G},\mathscr{M},\alpha)$ over $G$ is a bundle gerbe $\mathcal{G}$ on $G$ with
\begin{enumerate}
    \item an equivalence of gerbes (an invertible \emph{gerbe bimodule})
    \be\label{eq:scrm} \mathscr{M}: p_1^* \CG \otimes  p_2^* \CG  \to m^*\CG \ee
    over $G\times G$ (cf.~\ref{eq:twisted_convolution_product});
    \item an isomorphism $\alpha$ implementing associativity of group multiplication, namely
    \be
    \begin{tikzcd}
\CG_1\otimes \CG_2\otimes \CG_3 \arrow[d, "{{\rm id }\otimes \mathscr{M}_{2,3}}"'] \arrow[rr, "{\mathscr{M}_{1,2}\otimes {\rm id}}"] &  & \CG_{12}\otimes \CG_3 \arrow[d, "{\mathscr{M}_{12,3}}"] \arrow[lld, "{\alpha_{1,2,3}}", Rightarrow] \\
\CG_{1}\otimes \CG_{23} \arrow[rr, "{\mathscr{M}_{1,23}}"']                                                                          &  & \CG_{123}                                                                                                      
\end{tikzcd}\ee
where we have abbreviated $p_{k}^*\CG$ as $\CG_k$, $m_{i,j}^*\CG$ as $\CG_{ij}$ and $m^*_{1,2}\circ m_{12,3}^*\CG \cong  m_{2,3}^*\circ m_{1,23}^*\CG$ as $\CG_{123}$;
    \item the isomorphism shall satisfy the pentagon identity
    \be \ba &[{\rm id} \circ ({\rm id} \circ \alpha_{1,2,3})] \circ [ \alpha_{1,23,4}\circ {\rm id} ]\circ[{\rm id}\circ (\alpha_{1,2,3}\otimes {\rm id})] \\ = &[\alpha_{12,3,4}\circ {\rm id}] \circ [\alpha_{1,2,34}\circ{\rm id} ] \ea\ee
    which presents two identical ways to transform
    \be
    \mathscr{M}_{123,4} \circ (\mathscr{M}_{12,3}\otimes {\rm id}_{\CG_4}) \circ (\mathscr{M}_{1,2}\otimes {\rm id}_{\CG_3}\otimes {\rm id}_{\CG_4} )
    \ee
    into
    \be
    \mathscr{M}_{1,234} \circ ({\rm id}_{\CG_1}\otimes \mathscr{M}_{2,34}) \circ ({\rm id}_{\CG_1}\otimes {\rm id}_{\CG_2} \otimes \mathscr{M}_{3,4} ) \ .
    \ee
\end{enumerate}
\end{definition}
A conscious reader would immediately note that the $(\mathcal{G},\mathscr{M},\alpha)$ data satisfy almost identical conditions as $(K,\theta)$. An even better surprise is given by Proposition 5.2 of~\cite{Carey:2004xt}:
\begin{proposition}{Classification of multiplicative bundle gerbe}{}
    Let $G$ be a compact, connected Lie group. Then there is an isomorphism between $H^4(B G , \mathbb{Z})$ and the space of isomorphism classes of multiplicative bundle gerbes $(\mathcal{G},\mathscr{M},\alpha)$ on $G$.
\end{proposition}
Therefore we can obtain the tensor product structure on $\Sky^\tau(G)$ by first specifying the multiplicative bundle gerbe $\CG$ corresponding to $\tau\in H^4(BG,\bbZ)$, and then for every $X,Y\in {\rm Obj}(\Sky^\tau(G))$~\cite{weis2022a} (cf.~(\ref{eq:twisted_convolution_product})),
\be\label{eq:convolution_in_G}  X * Y = m_*(\mathscr{M} \otimes p_1^*X \otimes p_2^*Y) \ ,  \ee
exactly analogous to
\be  \bbC_x * \bbC_y = K_{x,y} \otimes \bbC_{xy} \ . \ee
We see that the gerbe bimodule $\mathscr{M}$ in~(\ref{eq:convolution_in_G}) is a more precise way to describe the convolution tensor product of $\Sky^\tau(G)$ than~(\ref{eq:twisted_convolution_product}), which by now should be understood as a ``bundle approximation'' of~(\ref{eq:convolution_in_G}), the precise meaning of which will be addressed in a moment.

A multiplicative bundle gerbe over $G$ is viewed as a bundle gerbe on $G$ endowed with additional multiplicative structure. This raises the question of how its Dixmier–Douady class in $H^{3}(G,\mathbb{Z})$ is reflected in the cohomology of the classifying space. In fact, one recovers in this way the suspension homomorphism $s:H^{4}(BG,\mathbb{Z}) \to H^{3}(G,\mathbb{Z})$, whose inverse is referred to as the transgression~\cite{Borel1955Topology, Dijkgraaf:1989pz} in the study of characteristic classes. Thus, multiplicative bundle gerbes are bundle gerbes whose Dixmier–Douady classes are in the image of $s$. For compact, simple and simply-connected $G$, $s$ is an isomorphism, and every bundle gerbe is multiplicative in this case.

To summarize, the anomaly information $\tau\in H^4(BG,\bbZ)$ uniquely defines a multiplicative bundle gerbe $(\CG,\mathscr{M},\alpha)$ where $\mathscr{M}$ determines the fusion in $\Sky^\tau(G)$. I.e., we define $\Sky^\tau(G)$ in exactly the same manner as Definition~\ref{def:symcatu1}, with $U(1)$ replaced by $G$ and~(\ref{eq:SkyU1_convolution}) replaced by~(\ref{eq:convolution_in_G}).

It will be illuminating to work out explicitly a more physically accessible description of the line bundle $K$ associated to $\tau \in H^4(BG, \mathbb{Z})$, or equivalently to $\mathscr{M}$ in $(\mathcal{G}, \mathscr{M}, \alpha)$. This is analogous to describing the first Chern class of a $U(1)$-bundle in terms of the field strength of a connection. One can then ask for a differential form description of $\tau$ in terms of multiplicative bundle gerbe connections, which is most naturally expressed in the local picture. As we explained in footnote~\ref{fn:gerbe}, a gerbe without connection is given by a collection $\{h_{i j k}\}$ on triple overlaps. To introduce a connection, one adds a connective structure $\{A_{i j} \in \Omega^1(\mathcal{U}_{i j})\}$ and 2-forms $\{B_{i} \in \Omega^2(\mathcal{U}_{i})\}$ such that
\begin{equation}
    B_j - B_i = d A_{i j} \quad \text{on} \;\, \mathcal{U}_{i j}\,, \quad  A_{i j} + A_{j k} + A_{k i} = d \,\text{log}\, h_{i j k}\quad \text{on} \;\, \mathcal{U}_{i j k}\,.
\end{equation}
The triple $(h, A, B)$ is the \v{C}ech data of a \emph{Deligne--Beilinson} cocycle~\cite{Deligne1971, Beilinson1985}, defining a class in $\check{H}^3(G)$ that models \emph{differential cohomology}~\cite{Cheeger1985,Hopkins:2002rd}. We now represent the differential refinement of $\tau$ by the tuple $(H, \rho)$ for certain $H\in \Omega^3(G)$ and $\rho\in \Omega^2(G\times G)$ which satisfy~\cite{bott1976rham, waldorf2010multiplicative, waldorf2016transgression}: 
\begin{equation}
    p_1^* H + p_2^* H = m^*H + d\rho
\end{equation}
for $m,p_1,p_2$ in~(\ref{eq:twisted_convolution_product}).
Given a trivialization of the bundle gerbe we have $H = dB$ on a local patch, i.e. $H$ is identified with the curvature of the gerbe connection $B$, hence the above equation becomes:
\begin{equation}\label{eq:trivilize_bundle_gerbe}
    p_1^*B + p_2^*B = m^*B + \rho + F
\end{equation}
with $dF = 0$. This establishes an invertible map from $\mathcal{I}_{p_1^*B}\otimes\mathcal{I}_{p_2^*B}$ to $\mathcal{I}_{m^*B}\otimes \mathcal{I}_{\rho}$, where $\mathcal{I}_{B}$ denotes trivial bundle gerbe with \emph{curving} $B\in \Omega^2(G)$ (globally defined connection $2$-form), which is formalized as an \emph{invertible bimodule} over $G\times G$ by Definition 1.3 of~\cite{waldorf2016transgression}. One can now view this map as an isomorphism from the trivial bundle gerbe $\mathcal{I}_{p_1^*B + p_2^*B}$ to $\mathcal{I}_{m^*B + \rho}$ on $G\times G$. By Proposition 1.1 of~\cite{waldorf2010multiplicative} (also see Section 3 of~\cite{waldorf2007more}), this invertible bimodule determines uniquely a line bundle over $G\times G$ with curvature $m^*B + \rho - p_1^*B - p_2^*B$, which is exactly $F$ in~(\ref{eq:trivilize_bundle_gerbe}). Therefore, we obtain a line bundle $K$ over $G\times G$ characterized by the ``de Rham data'', i.e. the curvature $F$ on each local patch of a trivialization. We see that this also addresses the precise meaning of~(\ref{eq:twisted_convolution_product}) being the ``bundle approximation'' of~(\ref{eq:convolution_in_G}) given a trivialization of the group manifold of $G$.

As a final remark, we note that since the topology of $G$ does matter, one has to distinguish $\Sky^{\tau^\delta}(G^\delta)$ from $\Sky^\tau(G)$, the former being the \emph{category of points} associated to the latter where $G^\delta$ is $G$ with discrete topology and $\tau^\delta$ the pull-back of $\tau$~\cite{weis2022a}. Since $G^\delta$ is discrete, $\tau^\delta$ is classified by~\cite{etingof2017tensor}:
\begin{equation}
    H^3(G^\delta, U(1)) \cong H^3(BG^\delta, U(1))
\end{equation}
which matches the computation using F-move of defect lines in~\cite{fiveguys_anomaly}. This in particular implies that $\Sky^{\tau^\delta}(G^\delta)$, rather than $\Sky^\tau(G)$, is the relevant category when studying flat gaugings of $G$. On the other hand, for continuous $G$ with ordinary topology, $\tau$ is classified by $H^4(BG,\mathbb{Z})$~\cite{Freed:2009qp}. The difference between the classifications of $\tau^\delta$ and of $\tau$ implies that the allowable 't Hooft anomalies of flat gauging can be different from the ones of a general dynamical gauging. However, since the homomorphism from $H^4(BG,\mathbb{Z})$ to $H^3(BG^\delta, U(1))$ is injective~\cite{Baez:2003yaq} for the group $G$ of our interests, we conclude that all anomalies of a general dynamical gauging can be detected by a suitably chosen flat gauging of $G$. In this work we do not emphasize the difference between $\Sky^{\tau^\delta}(G^\delta)$ and $\Sky^\tau(G)$ and have been focusing on the latter unless otherwise mentioned.

\subsection{Modular Data of $G=SU(2)$}
\label{sec:SU2}

The modular $S$-matrix for the Drinfeld center $Z(\textbf{Vec}_G)$ in the case of a finite group $G$ without twist takes the form
    \begin{equation}\label{Finite-group-S-matrix}
        S_{([a],\rho),([a'],\rho')} = \frac{1}{|G|} \sum_{\substack{g \in [a], h \in [a']\\gh=hg}} \chi_{\rho}^g(h)^* \chi_{\rho'}^{h}(g)^*,
    \end{equation}
where $\chi^g_{\rho}$ denotes the character of the representation $\rho$ transformed under the centralizer group $C_G(g)$. In the following, we assume the formula extends to the case of Lie groups to derive the $S$-matrix for $SU(2)$.

It is convenient to use the Euler-angle parametrization, where each $g\in SU(2)$ can be written as
    \begin{equation}
        g= g_{\phi} a_{\theta} g_{\psi}\,,
    \end{equation}
with
    \begin{equation}
        g_{\phi} = \left(\begin{array}{cc}
            e^{\frac{i\phi}{2}} & 0 \\
            0 & e^{-\frac{i\phi}{2}} 
        \end{array} \right) \,,\quad a_{\theta}=\left(\begin{array}{cc}
            \cos \frac{\theta}{2} & -\sin \frac{\theta}{2} \\
             \sin \frac{\theta}{2} & \cos \frac{\theta}{2}
        \end{array} \right)\,,
    \end{equation}
and
    \begin{equation}
        0\leq \theta \leq \pi\,,\quad 0 \leq \phi \leq 2\pi\,, \quad -2\pi \leq \psi \leq 2\pi\,.
    \end{equation}
The conjugacy class $\textrm{Cl}(SU(2))$ can be represented by the diagonal element $g_r$ with $r\in[0,2\pi]$ and we have
    \begin{equation}
        [g_r]:=\left\{x g_r x^{-1}|x \in SU(2) \right\}\,,\quad 0\leq r \leq 2\pi\,,
    \end{equation}
which consists of the elements
    \begin{equation}
        g(r,\theta,\phi) := g_{\phi} a_{\theta}g_ra^{-1}_{\theta} g^{-1}_{\phi}\,,
    \end{equation}
and they satisfy $g^{-1}(r,\theta,\phi) = g(-r,\theta,\phi) = g(r,\pi-\theta,\phi + \pi)$. We can define a left-invariant metric on $SU(2)$
    \begin{equation}
        \left(\textrm{Vol}(S^3)\right)^{\frac{2}{3}} g_{ij} = -\frac{1}{2}\textrm{Tr} \left(g^{-1} \partial_i g g^{-1} \partial_j g \right)\,,\quad (i,j=r,\theta,\phi)
    \end{equation}
where $\textrm{Vol}(S^3) = 2\pi^2$ is the volume of a unit 3-sphere. We then have
    \begin{equation}\label{SU(2)-Haar-measure}
        (2\pi^2)^{\frac{2}{3}}ds^2 = \frac{1}{4} d r^2 + \sin^2 \frac{r}{2} \left(d \theta^2 + \sin^2 \theta  d \varphi^2\right)\,.
    \end{equation}
The volume of $SU(2)$ is normalized to be one. An induced measure on the conjugacy class can be obtained by integrating $\theta,\phi$ out
    \begin{equation}\label{Conjugacy-class-measure}
        d s^2_{\textrm{Conj}} = \frac{1}{\pi^2}\sin^4 \frac{r}{2} dr^2\,. 
    \end{equation}

For $0<r<2\pi$, the centralizer $C_{SU(2)}(r,\theta,\phi)$ of $g(r,\theta,\phi)$ is the $U(1)$ parametrized by
    \begin{equation}
        C_{SU(2)}(r,\theta,\phi) = \left\{g(\alpha,\theta,\phi) | \alpha\in [-2\pi,2\pi) \right\}\,,
    \end{equation}
and the irreducible representations are labeled by the $U(1)$ charge $n\in \mathbb{Z}$. The character of the centralizer group is
    \begin{equation}
        \chi^{r}_{n} (\alpha) \equiv \chi^{g(r,\theta,\phi)}_{n}(g(\alpha,\theta,\phi)) = e^{\frac{i}{2}n \alpha}\,.
    \end{equation}
For $r=0,2\pi$, the centralizer is $SU(2)$ itself and the irreducible representations are labeled by the non-negative $j$ with the character
    \begin{equation}
        \chi_j(g(r,\theta,\phi)) = \frac{\sin\frac{j+1}{2}\beta}{\sin \frac{\beta}{2}}\,.\quad (j=0,1,2\cdots)
    \end{equation}

We now proceed to evaluate the $S$-matrix $S_{(r,\rho),(r',\rho')}$ in \eqref{Finite-group-S-matrix} for $G=SU(2)$. Since the appropriate way to formulate the summation in \eqref{Finite-group-S-matrix} as an integral is unclear to us, we will neglect the possible overall coefficients of the $S$-matrix in the following. The $S$-matrix depends on the centralizer of the group elements in $[g_r]$ and $[g_{r'}]$ and there exists four distinct cases
\begin{itemize}
    \item When $0<r,r'<2\pi$, then centralizer for both $[g_r]$ and $[g_{r'}]$ are $U(1)$. For a fixed $g(r,\theta,\phi) \in [g_r]$, the intersection between its centralizer $C_{SU(2)}(r,\theta,\phi)$ with the conjugacy class $[g_{r'}]$ is
        \begin{equation}
   C_{SU(2)}(r,\theta,\phi)\cap [g_{r'}] = \{g(r',\theta,\phi),g(-r',\theta,\phi) \}
        \end{equation}
where we used $g(-r',\theta,\pi)=g(r',\pi-\theta,\phi+\pi) \in [g_{r'}]$. We then have
    \begin{equation}
        S_{(r,n),(r',n')} = e^{-\frac{i}{2}(nr'+n'r)} + e^{\frac{i}{2}(nr'+n'r)} = 2\cos\frac{1}{2}(nr'+n'r)\,.
    \end{equation}
    \item When $0<r<2\pi$ and $r'=0,2\pi$ the conjugacy class of $r'$ consists only a single element, $1$ or $-1$. It trivially commutes with $g(r,\theta,\phi)\in [g_r]$ and we can write 
        \begin{equation}
            S_{(r,n),(0,j)} = \frac{\sin \frac{j+1}{2} r}{\sin \frac{r}{2}}\,,\quad S_{(r,n),(2\pi,j)} = (-1)^{n} \frac{\sin \frac{j+1}{2} r}{\sin \frac{r}{2}}\,.
        \end{equation}
    \item When $0<r'<2\pi$ and $r=0,2\pi$, we similarly have
        \begin{equation}
            S_{(0,j),(r,n)} = \frac{\sin \frac{j+1}{2} r}{\sin \frac{r}{2}}\,,\quad S_{(\pi,j),(r,n)} = (-1)^n\frac{\sin \frac{j+1}{2} r}{\sin \frac{r}{2}}\,.
        \end{equation}
    \item When both $r,r'=0,2\pi$, we have
        \begin{equation}
            \begin{gathered}
            S_{(0,j),(0,j')} = (j+1)(j'+1)\,,\quad S_{(2\pi,j),(0,j')} = (-1)^{j'}(j+1)(j'+1)\,,\\
            S_{(0,j),(2\pi,j')} = (-1)^{j}(j+1)(j'+1)\,,\quad S_{(2\pi,j),(2\pi,j')} = (-1)^{j+j'}(j+1)(j'+1)\,.
            \end{gathered}
        \end{equation}
\end{itemize}
In Appendix~\ref{sec:appE}, we check that the $S$-matrices introduced above satisfy the following properties
    \begin{equation}\label{SU(2)-S-properties}
        \begin{gathered}
            \frac{1}{4\pi}\int_{0}^{2\pi} d r' \sum_{n'}S_{(r,n),(r',n')}S_{(r',n'),(r'',n'')} = 2\pi \delta_{2\pi}(\frac{1}{2}r - \frac{1}{2}r'') \delta_{n,n''}\,,\\
            \frac{1}{4\pi}\int_0^{2\pi}  dr' \sum_{n'} S_{(0,j),(r',n')} S_{(r',n'),(r,n)}=0\,,\\
            \int_0^{2\pi}  (\frac{1}{\pi} \sin^2 \frac{r}{2})d r \sum_n S_{(0,j),(r,n)} S_{(r,n),(0,j')} = 2\pi \delta_{2\pi}(0) \delta_{j,j'}\,,\\
            \int_0^{2\pi}  (\frac{1}{\pi} \sin^2 \frac{r}{2})d r \sum_n S_{(2\pi,j),(r,n)} S_{(r,n),(0,j')} = 0\,,
        \end{gathered}
    \end{equation}
where we assume $0<r,r'<2\pi$ and $2\pi \delta_{2\pi}(0) = \sum_n 1=\sum_n e^{2\pi in \times 0}$. Notice that in the third and fourth lines, we add the measure of conjugacy class \eqref{Conjugacy-class-measure} while in the first and second lines we did not. The reason is that, for $S_{(0,j),(r,n)}$ (or $S_{(2\pi,j),(r,n)}$) and $S_{(r,n),(0,j)}$ the centralizers of $g_0=1$ and $g_{2\pi}=-1$ are the whole $SU(2)$, therefore the summation over conjugacy class $r$ should be considered as reduced from the measure of $SU(2)$. One the other hand, for $S_{(r,n),(r',n')}$ and $S_{(r',n'),(r'',n'')}$ the centralizer of $g_r$ and $g_{r''}$ are $U(1)$ so that $g_{r'} \in U(1)$ is reduced from the measure of $U(1)$.

Finally, let us examine the fusion coefficient by naively applying the Verlinde formula
    \begin{equation}\label{SU(2)-fusion-generic}
        N_{(r_1,n_1),(r_2,n_2)}^{(r_3,n_3)} = \frac{1}{4\pi}\int_{0}^{2\pi} dr \sum_{n} \frac{S_{(r_1,n_1),(r,n)}S_{(r_2,n_2),(r,n)}S^*_{(r_3,n_3),(r,n)}}{S_{(0,0),(r,n)}}\,.
    \end{equation}
where, for simplicity, we consider the most generic case where $0<r_1,r_2,r_3<2\pi$ and we choose $r_1>r_2$. Then we have
    \begin{equation}
        N_{(r_1,n_1),(r_2,n_2)}^{(r_3,n_3)} = 2\pi\delta_{2\pi}(\frac{1}{2}r_1+\frac{1}{2}r_2-\frac{1}{2}r_3) \delta_{n_3,n_1+n_2} + 2\pi\delta_{2\pi}(-\frac{1}{2}r_1+\frac{1}{2}r_2+\frac{1}{2}r_3) \delta_{n_3,n_1-n_2}\,,
    \end{equation}
where the proof is in Appendix~\ref{sec:appE}. It implies the fusion rule
    \begin{equation}
        W_{(r_2,n_2)} \times W_{(r_1,n_1)} =\frac{1}{4\pi} \int_0^{2\pi} dr_3 \sum_{n_3} N_{(r_1,n_1),(r_2,n_2)}^{(r_3,n_3)} W_{(r_3,n_3)}=W_{(\overline{r_1+r_2},n_1+n_2)} +W_{(r_1-r_2,n_1-n_2)}\,,
 \end{equation}
where $\overline{r_1+r_2}=\min(r_1+r_2,4\pi -r_1-r_2)$. Notice that it is different from the fusion rule given in \cite{Koornwinder:1998xg,Bais:1998yn}. As another example, let us set $r_1=r_2=0$. If $r_3\neq 0$, we have
    \begin{equation}
        N_{(0,n_1),(0,n_2)}^{(r_3,n_3)}=\frac{1}{4\pi} \int_0^{2\pi} dr \sum_n \frac{\sin \frac{n_1+1}{2}r}{\sin \frac{r}{2}} \frac{\sin \frac{n_2+1}{2}r}{\sin \frac{r}{2}} 2\cos\frac{1}{2}(n_3r+nr_3)=0\,,
    \end{equation}
where the summation over $n$ gives a delta function $\delta_{2\pi}(\frac{1}{2}r_3)$, which is zero given $r_3\in[0,2\pi]$ and $r_3 \neq 0$. On the other hand, if $r_3=0$, we have
    \begin{equation}\label{SU(2)-Fusion-rzero}
        N_{(0,n_1),(0,n_2)}^{(0,n_3)} = \int_0^{2\pi} \frac{1}{\pi} \sin^2 \frac{r}{2} dr\sum_n \frac{\sin \frac{n_1+1}{2}r}{\sin \frac{r}{2}} \frac{\sin \frac{n_2+1}{2}r}{\sin \frac{r}{2}} \frac{\sin \frac{n_3+1}{2}r}{\sin \frac{r}{2}}\,,
    \end{equation}
which can be evaluated directly and get
    \begin{equation}\label{SU(2)-Fusion-rzero-2}
        N_{(0,n_1),(0,n_2)}^{(0,n_3)} = 2\pi \delta_{2\pi}(0)\times \left\{ \begin{array}{l}
        1\quad n_3=|n_1-n_2|\,, |n_1-n_2|+2\,,\cdots\,, |n_1+n_2|\\
        0\, \quad \textrm{otherwise}
        \end{array}\right. \,.
    \end{equation}
It is the fusion coefficient among the irreducible representations of $SU(2)$. In fact, \eqref{SU(2)-Fusion-rzero} can be written as
    \begin{equation}
        N_{(0,n_1),(0,n_2)}^{(0,n_3)} = 2\pi \delta_{2\pi}(0)\int dg \chi_{n_1}(g) \chi_{n_2}(g) \chi^*_{n_3}(g)\,,
    \end{equation}
where $dg$ is the Haar measure \eqref{SU(2)-Haar-measure} and $\chi_n$ is the character of $SU(2)$. Equation \eqref{SU(2)-Fusion-rzero-2} then follows by the orthogonality relation of character.

\subsection{Lagrangian Algebra}
\label{sec:SU(2)-Lag}

We present some examples of Lagrangian algebras $\mc{L}$ for $G=SU(2)$ in the anomaly-free case. The requirement is that the lines $(r,n)$ appearing in $\mc{L}$ have a trivial mutual braiding with each other, i.e. the matrix elements $S_{(r,n),(r'n')}$ for all $(r,n),(r',n')\in\mc{L}$ are real and non-negative. We are not careful about the coefficients in $\mc{L}$ in this section, and the results are rigorous at least in the sense of weak gauging in \cite{Choi:2023xjw}.

We first have the Lagrangian subalgebra
\be
\mc{L}_{\rm Dir}=\bigoplus_n W_{(0,n)}
\ee
corresponding to the 2d theory with $SU(2)$ global symmetry. There is also the case of
\be
\mc{L}_{\rm Neu}=\bigoplus_r W_{(r,0)}
\ee
corresponding to the case where $SU(2)$ global symmetry was gauged with a flat connection, and there is a resulting dual (quantum) Rep$(SU(2))$ global symmetry.

Besides these, we also have the mixed boundary condition describing the gauging of the center $\mb{Z}_2\subset SU(2)$:
\be
\mc{L}_{SO(3)}=\bigoplus_{n=0}^\infty \left(W_{(0,2n)}\oplus W_{(2\pi,2n)}\right)\,,
\ee
giving a $\mb{Z}_2\times SO(3)$ 0-form symmetry with no anomaly.

Furthermore, we find a new class of boundary conditions which is not discussed in \cite{Jia:2025jmn}, which is for $q\in\mb{Z}$, $q>2$:
\be
\mc{L}_q=\bigoplus_m\bigoplus_{n=0}^{q-1}W_{(\frac{2\pi}{q}n,qm)}\,.
\ee
This boundary condition exactly corresponds to the gauging of a non-normal subgroup $\mb{Z}_q$ of the $SU(2)$ global symmetry, whose elements in $2\times 2$ matrices are
\be
\mb{Z}_q= \Big\{ \bp e^{\frac{2\pi in}{q}} & 0\\0 & e^{-\frac{2\pi in}{q}}\ep \Big\} \, .
\ee
The boundary theory would have a $\mb{Z}_q$ dual symmetry, as well as a non-invertible symmetry, following \cite{Tachikawa:2017gyf}. The non-invertible symmetry is described by a fusion category whose objects are $([g],\rho)$ where $[g]\in \mb{Z}_q\setminus SU(2)/\mb{Z}_q$ is a double coset and $\rho$ is an irreducible representation of $\mb{Z}_q\cup g\mb{Z}_q g^{-1}$~\cite{ostrik2003module}. Such ``coset'' symmetry~\cite{Hsin:2024aqb,Hsin:2025ria} is an example among the vast unexplored set of continuous non-invertible symmetries~\cite{GarciaEtxebarria:2022jky,Antinucci:2022eat,Damia:2023gtc,Delmastro:2025ksn}.

We would also like to comment that there seems to be no Lagrangian algebra corresponding to the gauging of $D$ or $E$-type non-abelian, non-normal subgroups of $SU(2)$. We would postpone the investigation of this issue to future works.

\section{The Issue of Finiteness}
\label{sec:Finiteness}

We have noted that in the computation of the Lagrangian algebra, a summation over infinite number of objects was encountered, e.g. in~(\ref{eq:LDir}). This poses the issue that if the constraint of the finiteness of the support can be relaxed~\footnote{Actually, it was pointed out in~\cite{Henriques:2015xxa} that though $\Sky^\tau(G)$ for non-abelian $G$ still makes sense, its center is too small to be interesting at all. Therefore, a generalization of $\Sky$ is necessarily required as long as a sensible center is needed.}. As mentioned in Section~\ref{sec:Gauging_Cyclic}, this amounts to finding a suitable $\mathscr{C}(G)$ of which $\Sky(G)$ is a subcategory. We will see that on our way of resolving this technical issue, a promising proposal of constructing symmetry category of general $G$ emerges. For simplicity we will focus on the non-anomalous case. We will also focus on the case where $G$ is an affine algebraic group, whose mathematical structures are to our knowledge best-investigated. We emphasize that e.g. the underlying group manifold of $U(1)$ or $SU(2)$ are real hence do not fall into the category of affine algebraic groups. We will comment on that issue later in this section.

We start with the fact that for finite $G$, $\Vec_G$ is actually equivalent to the category of quasi-coherent sheaves of finite dimensional vector spaces with the tensor product of convolution of sheaves~\cite{gelaki2015module}~\footnote{We assume certain finiteness condition to force a quasi-coherent sheaf to be coherent on finite group.}. We will denote this category over a (group) scheme $X$ by $\QC(X)$ with unit object $\mathbb{C}_e$, not to be confused with $\QCoh(X)$, the category of quasi-coherent sheaves of finite dimensional vector spaces with ordinary tensor product whose unit object is $\mathcal{O}_X$.~\footnote{Hence $\QCoh$ and $\QC$ have the same set of objects, but are equipped with different tensor product operations.} In the notation of~\cite{Ben-Zvi:2008vtm, ben2012morita}, $\QC(G)$ can be written as:
\begin{equation}
    \QC(G) \simeq \QC(\pt\times_{BG}\pt)
\end{equation}
using $G = \pt\times_{BG}\pt$~\cite{ben2012morita}. Theorem 1.9 of~\cite{Ben-Zvi:2008vtm} then implies:
\begin{equation}
    Z(\QC(G)) \simeq Z(\QC(\pt\times_{BG}\pt)) \simeq \QCoh(\mathcal{L}BG)
\end{equation}
where $\mathcal{L}BG$ is the free loop space of $BG$. Further applying Theorem 1.7 of~\cite{Ben-Zvi:2008vtm} with $X = BG$ and using the fact that $\QCoh(BG) \simeq \Rep_G$~\cite{Drinfeld2013, gaitsgory2019study}, we have:
\begin{equation}
    Z(\QC(G)) \simeq \QCoh(\mathcal{L}BG) \simeq Z(\Rep_G)\,.
\end{equation}
The above categorical equivalence is readily seen to be a straightforward generalization of $Z(\Vec_G) \simeq Z(\Rep_G)$ for finite $G$ to any affine algebraic group $G$. The category $\QC$ has the advantage over $\Sky$ that any (possibly infinite) family of objects in $\QC$ has a direct sum also in $\QC$. Therefore,~(\ref{eq:LDir}) is immediately sensible once we replace $\Sky$ by $\QC$. Since $\Sky(G)$ is a subcategory of $\QC(G)$, one may view $\QC(G)$ as the desired generalization, i.e. the sought-after $\mathscr{C}(G)$, in which the constraint of the finiteness of the support can be relaxed. Therefore, it is tempting to suggest that the symmetry category of a non-anomalous $G$-symmetry is $\QC(G)$. For anomalous $G$-symmetry, we note that the twisted convolution tensor product~(\ref{eq:twisted_convolution_product}) can also be defined on objects of $\QC$~\footnote{More precisely, the bundle gerbe $K$ is defined when both $X$ and $Y$ are simple objects in $\QC(G)$.}, thus making it $\QC^\tau(G)$ for $\tau \in H^4(BG,\mathbb{Z})$. Therefore, it is natural to expect that the symmetry category of anomalous $G$-symmetry is $\QC^\tau(G)$.

There is a crucial subtlety mentioned earlier that has not been discussed yet, that is the fact that many (Lie) groups encountered in Nature are not affine algebraic groups. This technicality is subtle but can potentially be resolved by considering the Lie algebra $\mathfrak{g}$ and the corresponding formal Lie group, denoted by $\exp(\mathfrak{g})$. One can show that $\QCoh(B\exp(\mathfrak{g}))$ is canonically equivalent to the category $\mathfrak{g}$-mod (see Chapter 7, Section 5 of~\cite{gaitsgory2017study}). In this sense $\QCoh(BG)$ and all $\QCoh(\mathcal{L}^nBG)$ are still well-defined. For $\QC$, we note that one can either consider $\QC(G_{\mathbb{C}})$ which falls within our previous discussion, or one may have to consider quasi-coherent and coherent sheaves on $C^\infty$-schemes developed in~\cite{joyce2012d, joyce2014introduction}. On physical ground it is natural to expect that the discussions for affine algebraic groups can be be generalized with no essential modification to $\QCoh$ (and $\QC$) on $C^\infty$-schemes, though a full-fledged mathematical proof (that whether the main results of~\cite{Ben-Zvi:2008vtm, ben2012morita} can be applied to $C^\infty$-schemes) is beyond the scope of this work.

\section{Discussions and Outlook}
\label{sec:discussions}

In this work we generalize the construction of the symmetry category and the corresponding SymTFT to continuous symmetry group, in particular for $G = U(1)^n$ and for compact connected non-abelian Lie group $G$. For this, it is crucial to realize that the proper generalization of $\Vec_G^\tau$ for finite $G$ is $\Sky^\tau(G)$ for $\tau\in H^4(BG,\mathbb{Z})$, whose corresponding SymTFT, i.e. Drinfeld center, can be computed in the manner proposed in~\cite{Freed:2009qp} as we reviewed in section~\ref{sec:Z(Sky(G))}. A subtle but important technicality is that in $\Sky^\tau(G)$ only the objects with finite support over $G$ are allowed, in which sense $\Sky^\tau(G)$ becomes a minimal generalization of $\Vec_G^\tau$ for finite $G$ for which the finiteness of the support is automatically guaranteed.

As it was shown that the SymTFT of (non-anomalous) continuous flavor $G$-symmetry is a $BF$-theory~\cite{Jia:2025jmn}, it will be interesting to see how the categorical data of $\Sky(G)$ is encoded in $BF$-theory, thereby providing a concrete realization of $\Sky(G)$ at the level of the action principle. This, for 2D QFT $\mathcal{T}_G$, amounts to viewing the 3D $BF$-theory with gauge group $G$ as a 0-1-2-3 fully extended topological field theory~\cite{Baez:1995xq, Lurie:2009keu} and assigning $\Sky(G)$ to a point of a 3D manifold in a proper way.

Certainly, it is also worth exploring if one can generalize $\Sky^\tau(G)$ to $\QC^\tau(G)$ to resolve the finiteness issue as discussed in Section~\ref{sec:Finiteness}. This clearly involves generalizing the notion of quasi-coherent to a $C^\infty$ setting, and this is an interesting mathematical problem to pursue.

A potentially very concrete follow-up would be to classify all Lagrangian algebras of $\Sky^\tau(SU(2))$ for any $\tau \in H^4(BSU(2),\mathbb{Z})$, which we left open in the work. Such a classification would enable us to achieve a deeper understanding of the gauging and the phase structure of $\mathcal{T}^\tau_{SU(2)}$. Certainly, a further generalization to arbitrary compact $G$ would be a huge step towards understanding the gaugings and the phases of theories with continuous global symmetries.

An even wilder dream is to push the machinery of manifold tensor category to its limit, in the sense that one can now study $\Sky^\tau(G)$ for any group manifold $G$ or even groupoid as long as a sensible monoidal tensor structure can be defined. This may put the study of both internal and spacetime symmetries in a uniform setting. The crucial part is to classify all consistent $\tau$'s for general $G$. This will enable us to explore the anomalies of a very general class of global symmetries.

Another direction is to consider the $\mathbb{Z}_2$-graded tensor category $\mathbf{sSky}^{\tau/2}(U(1))$ whose objects are skyscraper sheaf of finite-dimensional complex super vector spaces over $G$ with finite support. It can be twisted by a class $\frac{1}{2}\tau$ which is half of the generator of $H^4(BU(1),\mathbb{Z})$, and the pair $(K,\theta)$ are also lifted to hermitian $\mathbb{Z}_2$-graded line bundles. The existence of a half class requires a spin structure and thus $\mathbf{sSky}^{\tau/2}(U(1))$ defines a symmetry category of $U(1)$ fermionic theory. It is also worthwhile to study the Drinfeld center $Z(\mathbf{sSky}^{\tau/2}(U(1)))$, which may provide a realization of a $(2+1)d$ fermionic SymTFT associated with $U(1)$ symmetry.

\section*{Acknowledgement}
We thank Theo Johnson-Freyd, Gen Yue and Yunqin Zheng for various helpful discussions. YZ thanks Yuji Tachikawa for valuable discussions and explanations on various aspects of anomalies. RL and YNW are supported by National Natural Science Foundation of China under Grant No. 12175004, No. 12422503 and by Young Elite Scientists Sponsorship Program by CAST (2024QNRC001). JT is supported by National Natural Science Foundation of China under Grant No. 12405085 and by the Natural Science Foundation of Shanghai (Grant No. 24ZR1419300). JT would also like to thank Ying Zhang for her love and support.
YZ is supported by WPI Initiative, MEXT, Japan at Kavli IPMU, the University of Tokyo and by National Science Foundation of China (NSFC) under Grant No. 12305077. QJ is supported by National Research Foundation of Korea (NRF) Grant No. RS-2024-00405629 and Jang Young-Sil Fellow Program at the Korea Advanced Institute of Science and Technology.
This research was supported in part by Perimeter Institute for Theoretical Physics. Research at Perimeter Institute is supported by the Government of Canada through the Department of Innovation, Science and Economic Development and by the Province of Ontario through the Ministry of Colleges and Universities.

\appendix

\section{\texorpdfstring{$U(1)$}{U(1)} anomaly in 2d and invertible phase}
\label{sec:appA}
For $2d$ theory with spin structure, the $U(1)$ anomaly is characterized by the \emph{invertible phase}~\cite{Freed:2016rqq} ${\rm Inv}^3_{\rm spin}(BU(1))$ which fits into the short exact sequence
\begin{equation} 
    0 \longrightarrow  \text{Ext}_{\mathbb{Z}}(\Omega^{\rm spin}_{3}(BU(1)),\mathbb{Z})\longrightarrow  \text{Inv}^{3}_{\rm spin}(BU(1)) \longrightarrow \text{Hom}_{\mathbb{Z}}( \Omega^{\text{spin}}_{4}(BU(1)),\mathbb{Z})\longrightarrow 0 \,,
\end{equation}
where $\text{Ext}_{\mathbb{Z}}(\Omega^{\rm spin}_{3}(BU(1)),\mathbb{Z})= {\rm Hom}_{\mathbb{Z}}({\rm Tor}[\Omega_3^{\rm spin}(BU(1))], U(1))$ accounts for the torsional part of anomaly while ${\rm Hom}_{\mathbb{Z}}( \Omega^{\text{spin}}_4(BU(1)),\mathbb{Z}) = {\rm Hom}_{\mathbb{Z}}({\rm Free}[\Omega_4^{\rm spin}(BU(1))], \mathbb{Z})$ captures the free part.
For spin bordism group $\Omega_\bullet^{\rm spin}(X)$, there is a general decomposition
\begin{equation}
    \Omega_\bullet^{\rm spin}(X) = \Omega_\bullet^{\rm spin}(\rm pt.) \oplus \tilde{\Omega}_\bullet^{\rm spin}(X) \,,
\end{equation}
where the first factor is its value at the point and second one is called the \emph{reduced} bordism group. 
The relevant result of $\Omega_\bullet^{\rm spin}(BU(1))$ can be found in~\cite{Garcia-Etxebarria:2018ajm}.
\begin{equation}
    \begin{tabular}{c|cccccc}
  $p$ & $0$ & $1$ & $2$ & $3$ & $4$ & $5$\\ 
$\Omega_p^{\rm spin}(BU(1))$ & $\mathbb{Z}$ & $\mathbb{Z}$ & $\mathbb{Z}_2 \oplus \mathbb{Z} $ & 0 & $\mathbb{Z} \oplus \mathbb{Z}$ & $0$
\end{tabular}
\end{equation}
We see that $\text{Ext}_{\mathbb{Z}}(\Omega_3^{\rm spin}(BU(1)),\mathbb{Z})= {\rm Hom}_{\mathbb{Z}}({\rm Tor}[\Omega_3^{\rm spin}(BU(1))], U(1)) = 0$ and we have the isomorphism 
\begin{equation}
    {\rm Inv}^3_{\rm spin}(BU(1)) \cong {\rm Hom}_{\mathbb{Z}}( \Omega^{\text{spin}}_4(BU(1)),\mathbb{Z}) = {\rm Hom}_{\mathbb{Z}}({\rm Free}[\Omega_4^{\rm spin}(BU(1))], \mathbb{Z}) = \mathbb{Z} \oplus \mathbb{Z}\,,
\end{equation}
which exactly corresponds to the fact 
\begin{equation}
    \Omega_4^{\rm spin}(\rm pt) = \tilde\Omega_4^{\rm spin}(BU(1)) = \Z\,.
\end{equation}
The generator of $\Omega_4^{\rm spin}(\rm pt)$ is the $K3$ surface and it can be detected by the first Pontryagin class $p_1$ and we have
\begin{equation}
    \int_{K3} -\frac{1}{48}p_1 = 1\,.
\end{equation}
Thus, the dual basis of $\Omega_4^{\rm spin}(\rm pt)$ is given by $-\frac{1}{48}p_1$, corresponds to gravitational anomaly.

The reduced bordism group $\tilde\Omega_4^{\rm spin}(BU(1))$ captures the anomaly of internal $U(1)$ symmetry. It can be detected by $(c_1)^2$. However, the dual basis is $\frac{1}{2}(c_1)^2$, \emph{half} of the generator of $H^4(BU(1),\bbZ)$. The generator of $\tilde\Omega_4^{\rm spin}(BU(1))$ is $S^2 \times S^2$ with $U(1)$-bundle 
\begin{equation}
    c_1= a+b \in H^2(S^2\times S^2, \bbZ),
\end{equation}
where $a, b$ are the two $S^2$ generators. 
Then we recover the dual relation
\begin{equation}
    \int_{S^2\times S^2} \frac{1}{2}(c_1)^2 = 1\,.
\end{equation} The real reduction of $\frac{1}{2}(c_1)^2$ is the anomaly polynomial $\frac{1}{2}\frac{F}{2\pi} \wedge \frac{F}{2\pi}$, corresponds to Chern--Simons form $\frac{1}{2}\frac{A}{2\pi} \wedge \frac{F}{2\pi}$ at level 1. 

The ratio between these dual bases and characteristic classes has important physical consequences.  A notable example is the quantization condition for $U(1)$ Chern--Simons theory: 
the half-integral normalization associated with the spin bordism generator implies that odd level $U(1)$ Chern--Simons theories can be consistently defined only on spin three-manifolds $M_3$~\cite{Dijkgraaf:1989pz, Belov:2005ze}. From the bordism group result, one can always find a spin four manifold $N_4$ whose boundary is $M_3$ such that the $U(1)$ bundle and spin structure on $N_4$ extending those on $M_3$. 
Now, using the Stokes theorem, we try to define as follows: \begin{equation}
\text{``}\int_{M_3} \frac{1}{2} \frac{A}{2\pi}  \wedge \frac{F}{2\pi}\text{''} \,{:=}\,  \frac{1}{2}\int_{N_4}  \frac{F}{2\pi} \wedge \frac{F}{2\pi}\,.
\end{equation}
As $(\frac{F}{2\pi})^2$ integrates to even integer on closed spin four manifold, under the presence of spin structure, the quantity $\exp 2 \pi \,i \int_{M_3} \frac{1}{2} \frac{A}{2\pi}   \frac{F}{2\pi}$ is independent of the extension and hence well-defined.

If we denote by $\widetilde{{\rm Inv}^3}_{\rm spin}(BU(1))$ the reduced invertible phase to forget about the gravitational part, then the $\frac{1}{2}$ means the following isomorphisms 
\begin{equation}
    \begin{split}
        H^4(BU(1),\bbZ) &\longrightarrow \widetilde{{\rm Inv}^3}_{\rm spin}(BU(1)) \cong \text{Hom}(\tilde\Omega_4^{\rm spin}(BU(1),\mathbb{Z}) \\
        \mathbb{Z}\ni 1& \mapsto 2 \in \mathbb{Z} \,,
    \end{split}
\end{equation}
while for $SU(2)$ this is 
\begin{equation}
    \begin{split}
        H^4(BSU(2),\bbZ) &\longrightarrow \widetilde{{\rm Inv}^3}_{\rm spin}(BSU(2)) \\
        \mathbb{Z}\ni 1& \mapsto 1 \in \mathbb{Z} \,.
    \end{split}
\end{equation}
These isomorphisms between bosonic and fermionic anomalies offer an alternative perspective on the fact that odd level $U(1)$ Chern--Simons theory can only be 
defined on spin manifold, whereas odd level $SU(2)$ Chern--Simons theory does not require spin structure.

If we take a class $\tau \in  H^4(BU(1),\bbZ)$ as the input anomaly data, then we are always reaching even level $U(1)$ Chern--Simons invariant, and hence landing in bosonic theories independent of spin structure.

\section{Equivalent descriptions of the anomaly four-cocycle $\tau$.}
\label{sec:appB}
The cohomology ring $H^\bullet(BU(1),\mathbb{Z})$ is the polynomial ring $\mathbb{Z}[c_1]$, where $c_1$ is the (universal) first Chern--Class generating $H^2(BU(1),\mathbb{Z})$. A cohomology class $\tau \in H^4(BU(1),\mathbb{Z}) \cong \mathbb{Z}$ is simply specified by an integer $k \in \mathbb{Z}$ and we have $\tau = k (c_1)^2$, here $(c_1)^2$ is the generator of $H^4(BU(1),\mathbb{Z})$.
 Hence, $H^4(BU(1),\bbZ)$ can also be thought as the set of homogeneous quadratic forms 
  \begin{equation}
    \begin{split}
        q: \Pi=H_2(BU(1),\mathbb{Z})  &\longrightarrow \mathbb{Z}\\
         n \cdot  \kappa &\mapsto k  n^2 \times 1\,,
         \end{split}
    \end{equation}
   where $\kappa$ is the generator of $H_2(BU(1),\mathbb{Z})$. We can define a bilinear (or bi-additive) form 
    \begin{equation}
    \begin{split}
        \expval{\cdot,\cdot}: H_2(BU(1),\mathbb{Z}) \times H_2(BU(1),\mathbb{Z})  &\longrightarrow \mathbb{Z}\\
         (n \cdot  \kappa, m\cdot \kappa )&\mapsto 2 k  nm \times 1 \,,
         \end{split}
    \end{equation}
  as the polarization for each $q$, i. e. $\langle n \cdot  \kappa, m \cdot \kappa\rangle = q  (n\cdot  \kappa+ m \cdot \kappa) - q  (n\cdot  \kappa) - q  (m \cdot \kappa)$. This assignment is a 1-1 correspondence and we see that $\langle \pi, \pi \rangle = 2q(\pi) \in 2\mathbb{Z}$, $\forall \pi \in\Pi$. Hence, $\expval{\cdot,\cdot}$ is an even form. $q$ is also called the quadratic refinement of $\expval{\cdot,\cdot}$.
   Let us take a class $\tau = k \,(c_1)^ 2$ with $k \neq 0$, then the corresponding even form $\expval{\cdot,\cdot}$ is non-degenerate over $\mathbb{Q}$~\cite{Freed:2009qp} and induces an injective homomorphism (still denoted by $\tau$)
   \begin{equation}
   \begin{split}
       \tau:\Pi=H_2(BU(1),\mathbb{Z})  &\longrightarrow  H^2(BU(1),\mathbb{Z}) = \Lambda \\
                          n \cdot  \kappa        &  \mapsto  2kn \cdot c_1\,, 
       \end{split}
   \end{equation}
   and we recover $\langle n\cdot\kappa, m \cdot\kappa \rangle = \left(\tau(n\cdot\kappa)\right)(m \cdot\kappa) = 2k n c_1(m \cdot\kappa)= 2k n m$ as before. Here we used the duality between $c_1$ and $\kappa$: $c_1(\kappa) = 1$.

In the case of $G=SU(2)$, the first Chern--Class vanishes and the cohomology ring  $H^\bullet(BSU(2),{\mathbb{Z}})$ is isomorphic to $\mathbb{Z}[c_2]$ and $c_2$ is the second Chern--Class that generating $H^4(BSU(2),{\mathbb{Z}}) \cong \mathbb{Z}$. Let us fix a maximal torus in $SU(2)$ and denote it by $T \subset SU(2)$. The corresponding root lattice is $\Lambda_r = \mathbb{Z} \alpha$.
We have the standard identification (see for example the textbook~\cite{BrockerDieck1985} for a review) 
\begin{equation}
    \begin{aligned}
 H^2(BT,{\mathbb{Z}}) & \cong \Lambda_w  \cong \mathbb{Z}\lambda \quad \text{weight lattice}\,, \\ 
 H_2(BT,{\mathbb{Z}}) & \cong  \Lambda_r^\vee \cong \mathbb{Z}\alpha^\vee \quad \text{coroot lattice}.
    \end{aligned}
\end{equation}
The Weyl group $W= {\mathbb{Z}}_2$ acts on $\Lambda_r$ as $\alpha \mapsto - \alpha$, the action extends accordingly to $\Lambda_w$ and $\Lambda_r^\vee$. We have the standard normalization: $(\alpha,\alpha)_{\rm Killing} = 2$ and $\alpha^\vee(\lambda)=1$, here $\lambda$ is the fundamental weight. 
$H^4(BSU(2),{\mathbb{Z}})$ can be regarded as the Weyl-invariant subgroup $H^4(BT,{\mathbb{Z}})^W \subset H^4(BSU(2),{\mathbb{Z}})$~\cite{Henriques:2016ipo}, and hence in a similar manner to the abelian case
\begin{equation}
    \begin{aligned}
H^4(BSU(2),{\mathbb{Z}}) &\cong \left({\rm Sym}^2(\Lambda_w)\right)^W \quad \textrm{Weyl-invariant symmetric product }\\
&\cong \{ \textrm{Weyl-invariant symmetric biadditive even forms} :\Lambda_r^\vee \times \Lambda_r^\vee \to {\mathbb{Z}} \}\\
&\cong \{ \textrm{Weyl-invariant even homomorphisms } \tau: \Lambda_r^\vee \to \Lambda_w \} \,.
    \end{aligned}
\end{equation}
The above discussion is in fact quite general for any compact connected Lie groups and we focus on $SU(2)$ for simplicity. In this case, since the Weyl group acts by a multiplication of $-1$ on the generators of each lattice, the Weyl-invariance is automatically satisfied.

\section{Modular data of $Z(\textrm{Vec}^{k}_{\mathbb{Z}_N})$ and $N\rightarrow \infty$ limit}\label{sec:appC}
In this appendix, we will first derive the modular data of $Z(\textrm{Vec}^{\omega}_{\mathbb{Z}_N})$ following \cite{Chen:2024ulc} using the 3d level-$N$ BF theory with the $k$-twist
\begin{equation}
        S = \frac{N}{2\pi} \int \widetilde{A}\wedge d A  + \frac{2 k}{4\pi} \int A \wedge d A,
    \end{equation}
where $k=0,1,\cdots,N-1$. Next, we will consider the continuum limit $N\rightarrow \infty$.

The gauge transformations are
    \begin{equation}
        A \rightarrow A + d \Lambda,\quad \widetilde{A} \rightarrow \widetilde{A} + d \widetilde{\Lambda}\,,
    \end{equation}
and we can consider the Wilson loop operators $W[\Gamma]$ and $\widetilde{W}[\Gamma]$ for $A$ and $\widetilde{A}$
    \begin{equation}
        W[\Gamma] = \exp \left(i \oint_{\Gamma} A \right)\,, \quad \widetilde{W}[\Gamma] = \exp \left(i \oint_{\Gamma} \widetilde{A} \right)\,.
    \end{equation}
The S-matrix elements between two kinds of line operators can be obtained by computing the correlation function
    \begin{equation}
        \langle W[\Gamma] \widetilde{W}[\Gamma'] \rangle = \int \mathcal{D}A \mathcal{D}\widetilde{A} e^{i \frac{N}{2\pi} \int \widetilde{A}\wedge d A + i \frac{2 k}{4\pi} \int A \wedge d A + i \int \eta(\Gamma) \wedge A + i \int \eta(\Gamma') \wedge \widetilde{A}},
    \end{equation}
where $\eta(\Gamma)$ is a 2-form and is the Poincare dual of $\Gamma$. Using the relation
    \begin{equation}
        \int \eta(\Gamma) \wedge A = \int d \eta(D_{\Gamma}) \wedge A = \int \eta(D_{\Gamma}) \wedge d A,
    \end{equation}
where $D_{\Gamma}$ is the 2-surface bounded by $\Gamma$ and $d \eta(D_{\Gamma}) = \eta(\partial D_{\Gamma}) = \eta(\Gamma)$. The phase on the exponent can be written as
    \begin{align}
         &\frac{N}{2\pi} \int \widetilde{A}\wedge d A +  \frac{k}{2\pi} \int A \wedge d A +\int \eta(\Gamma) \wedge A  + \int \eta(\Gamma') \wedge \widetilde{A} \nonumber\\
         =& \frac{N}{2\pi} \int \left(\widetilde{A} + \frac{2\pi}{N} \eta(D_{\Gamma}) - \frac{4\pi k}{N^2}\eta(D_{\Gamma'})\right)\wedge d \left(A + \frac{2\pi}{N} \eta(D_{\Gamma'}) \right) \nonumber\\
         &+  \frac{k}{2\pi} \int \left(A + \frac{2\pi}{N} \eta(D_{\Gamma'}) \right) \wedge d \left(A + \frac{2\pi}{N} \eta(D_{\Gamma'}) \right)  \nonumber \\
         &- \frac{2\pi}{N} \int \eta(D_{\Gamma}) \wedge d \eta(D_{\Gamma'})+\frac{2\pi k}{N^2} \int \eta(D_{\Gamma'}) \wedge d \eta(D_{\Gamma'})
    \end{align}
which indicates,
    \begin{equation}
        \langle W[\Gamma] \widetilde{W}[\Gamma'] \rangle = \langle W[\Gamma] \rangle \langle \widetilde{W}[\Gamma'] \rangle e^{-\frac{2\pi i}{N} \textrm{Link}(\Gamma,\Gamma')},
    \end{equation}
where the last term gives the linking number between $\Gamma$ and $\Gamma'$. The topological spin of $\widetilde{W}[\Gamma]$ can be read as $e^{\frac{2\pi i k}{N^2}}$. Consider the insertion of two $\widetilde{W}$ Wilson loops as $\widetilde{W}[\Gamma]$ and $\widetilde{W}[\Gamma']$, one can similarly deduce that
    \begin{equation}
        \langle \widetilde{W}[\Gamma] \widetilde{W}[\Gamma'] \rangle = \langle \widetilde{W}[\Gamma] \rangle \langle \widetilde{W}[\Gamma'] \rangle e^{\frac{4\pi k i}{N^2} \textrm{Link}(\Gamma,\Gamma')}.      
    \end{equation}
Moreover, since $W^N$ link trivially with other line operators, one can identify
    \begin{equation}
    W^N[\Gamma] = 1.
    \end{equation}
On the other hand, $\widetilde{W}^N$ link trivially with $W$ operators but one has
    \begin{equation}
        \langle \widetilde{W}^N[\Gamma] \widetilde{W}[\Gamma'] \rangle = \langle \widetilde{W}^N[\Gamma] \rangle \langle \widetilde{W}[\Gamma'] \rangle e^{\frac{4\pi k i}{N} \textrm{Link}(\Gamma,\Gamma')},      
    \end{equation}
therefore one should identify
    \begin{equation}
        \widetilde{W}^N[\Gamma] = W^{-2k}[\Gamma].
    \end{equation}

Let us denote generic line operators using a pair of indices $\alpha,\tilde{\alpha} = 0,1,\cdots,N-1$ as
    \begin{equation}
        W_{(\alpha,\tilde{\alpha})} [\Gamma] = \exp \left(i \oint \alpha A + \tilde{\alpha} \widetilde{A} \right).
    \end{equation}
Each line operator has quantum dimension one so that the total quantum dimension is $N$. The elements of $S$ and $T$ matrices can be read from above as\footnote{The convention of $T$-matrices is different from that in \cite{Chen:2024ulc}.}
    \begin{equation}
        S_{(\alpha,\tilde{\alpha}),(\beta,\tilde{\beta})} = \frac{1}{N}\omega^{-\alpha \tilde{\beta} - \tilde{\alpha} \beta + \frac{2k}{N} \tilde{\alpha} \tilde{\beta} },\quad T_{(\alpha,\tilde{\alpha}),(\beta,\tilde{\beta})} = \omega^{\alpha \tilde{\alpha} - \frac{k}{N} \tilde{\alpha}^2} \delta_{\alpha,\beta} \delta_{\tilde{\alpha},\tilde{\beta}}
    \end{equation}
with $\omega = e^{\frac{2\pi i}{N}}$. The fusion coefficient is obtained via the Verlinde formula

    \begin{equation}
        \begin{split}
            N_{(\alpha,\tilde{\alpha}),(\beta,\tilde{\beta})}^{(\gamma,\tilde{\gamma})} =& \sum_{\sigma,\tilde{\sigma}} \frac{S_{(\sigma,\tilde{\sigma}),(\alpha,\tilde{\alpha})} S_{(\sigma,\tilde{\sigma}),(\beta,\tilde{\beta})} S_{(\sigma,\tilde{\sigma}),(\gamma,\tilde{\gamma})}^*}{S_{(0,0),(\sigma,\tilde{\sigma})}} \\
            =& \frac{1}{N^2} \sum_{\sigma,\tilde{\sigma}} \omega^{-\sigma \tilde{\alpha} - \tilde{\sigma} \alpha + \frac{2k}{N}\tilde{\sigma}\tilde{\alpha}} \omega^{-\sigma \tilde{\beta} - \tilde{\sigma} \beta + \frac{2k}{N}\tilde{\sigma}\tilde{\beta}} \omega^{\sigma \tilde{\gamma} + \tilde{\sigma} \gamma - \frac{2k}{N}\tilde{\sigma}\tilde{\gamma}}\\
            =&\frac{1}{N} \sum_{\tilde{\sigma}} \delta_{\tilde{\alpha}+\tilde{\beta}-\tilde{\gamma},0} \omega^{ -\tilde{\sigma} (\alpha+\beta-\gamma) + \frac{2k}{N}\tilde{\sigma}(\tilde{\alpha}+\tilde{\beta}-\tilde{\gamma})}
        \end{split}
    \end{equation}
where the delta function is defined mod $N$ such that
    \begin{equation}
        \delta_{k,0} = \left\{ \begin{array}{l}
            1 \quad k = 0\ \textrm{mod}\ N\\
            0 \quad \textrm{others}
        \end{array}\right. .
    \end{equation}
Notice that $-N<\tilde{\alpha}+\tilde{\beta}-\tilde{\gamma}<2N-1$, we can further sum over $\tilde{\sigma}$ and get
    \begin{equation}
        N_{(\alpha,\tilde{\alpha}),(\beta,\tilde{\beta})}^{(\gamma,\tilde{\gamma})} = \delta_{\gamma,\alpha+\beta - 2k \left(\frac{\tilde{\alpha} + \tilde{\beta} - \tilde{\gamma}}{N}\right)} \delta_{\tilde{\gamma},\tilde{\alpha}+\tilde{\beta}}\,.
    \end{equation}

In the continuum limit $N\rightarrow \infty$, we can replace
    \begin{equation}
        \alpha \rightarrow n\in \mathbb{Z}\,,\quad \frac{2\pi \tilde{\alpha}}{N}\rightarrow \theta_x \in [0,2\pi)\,,
    \end{equation}
and introduce $x=e^{i\theta_x}$. Then the modular $S$ and $T$ matrices for $Z(\Sky^{\tau}(U(1)))$ takes the form
    \begin{equation}
        \widetilde{S}_{(x,n),(y,n')} = \frac{1}{2\pi}e^{-i(n\theta_x + n'\theta_y - \frac{2k}{2\pi} \theta_x \theta_y)},\quad T_{(x,n),(y,n')} = e^{i (n\theta_x - \frac{k}{2\pi} \theta_x^2)}  \delta_{x,y}\delta_{n,n'},
    \end{equation}
where we normalize the $S$-matrix as $\widetilde{S} \equiv \lim_{N\rightarrow \infty} \frac{N}{2\pi} S$. The fusion coefficient can be obtained in the same limit as
    \begin{equation}
        N^{(z,n'')}_{(x,n),(y,n')}=  \delta_{xy,z}\delta_{n'',n+n'- 2k \left(\frac{\theta_x+\theta_y-\theta_z}{2\pi}\right)}\,.
    \end{equation}

\section{Properties of $S/T$-matrices of $Z(\mathbf{Sky}^{\tau}(U(1)))$} \label{sec:appD}

In this appendix, we will prove the various properties of $S/T$-matrices given in section \ref{sec:modular-U(1)}. We need the following identities throughout this section
    \begin{equation}
        \sum_{k\in\mathbb{Z}} e^{i k \alpha} = 2\pi \delta_{2\pi}(\alpha)\,, \quad \int_0^{2\pi} d\alpha e^{i k \alpha} = 2\pi \delta_{k,0}\,,
    \end{equation}
where 
    \begin{equation}
        \delta_{2\pi}(\alpha)= \sum_{n\in \mathbb{Z}} \delta(\alpha+2\pi n)\,.
    \end{equation}

\subsection*{Proof of unitarity of $S$ and $T$}
It is easy to check $S$ is unitary
    \begin{equation}
    \begin{split}
        &\int_0^{2\pi} d\theta_z \sum_{n''} S^*_{(x,n),(z,n'')} S_{(z,n''),(y,n')}\\
        =&\int_0^{2\pi} d\theta_z \sum_{n''} \left(\frac{1}{2\pi} e^{i(n \theta_z+n'' \theta_x-\frac{2k}{2\pi} \theta_x \theta_z)} \right)\left(\frac{1}{2\pi} e^{-i(n'' \theta_y+n' \theta_z-\frac{2k}{2\pi} \theta_y \theta_z)} \right) \\
        =&\frac{1}{2\pi}\int_0^{2\pi} d\theta_z\left(\frac{1}{2\pi}\sum_{n''}e^{ i n''(\theta_x-\theta_y)} \right) e^{i \theta_z (n-n'+\frac{2k}{2\pi}(\theta_y-\theta_x))}\\
        =&\frac{1}{2\pi}\int_0^{2\pi} d\theta_z \delta_{2\pi}(\theta_x-\theta_y) e^{i \theta_z (n-n'+\frac{2k}{2\pi}(\theta_y-\theta_x))}\\
        =&\delta_{x,y} \delta_{n,n'}\,,
    \end{split}
    \end{equation}
where we use the fact that $\theta\in[0,2\pi)$ so that $\delta_{2\pi}(\theta_x-\theta_y)$ will impose $\theta_x=\theta_y$. $T$ is also unitary since
    \begin{equation}
    \begin{split}
        &\int_0^{2\pi} d\theta_z \sum_{n''} T^*_{(x,n),(z,n'')} T_{(z,n''),(y,n')}\\
        =&\int_0^{2\pi} d\theta_z \sum_{n''} e^{-i(n\theta_x -\frac{k}{2\pi}\theta_x^2)}e^{i(n''\theta_z -\frac{k}{2\pi}\theta_z^2)} \delta_{n,n''} \delta_{n'',n'} \delta_{2\pi}(\theta_x-\theta_z) \delta_{2\pi}(\theta_z-\theta_y)\\
        =&\delta_{x,y} \delta_{n,n'} \,.
    \end{split}
    \end{equation}

\subsection*{Proof of $S^2 =(ST)^3=C$}
For $S^2$, one has similarly

    \begin{equation}
    \begin{split}
        &\int_0^{2\pi} d\theta_z \sum_{n''} S^*_{(x,n),(z,n'')} S_{(z,n''),(y,n')}\\
        =&\int_0^{2\pi} d\theta_z \sum_{n''} \left(\frac{1}{2\pi} e^{-i(n \theta_z+n'' \theta_x-\frac{2k}{2\pi} \theta_x \theta_z)} \right)\left(\frac{1}{2\pi} e^{-i(n'' \theta_y+n' \theta_z-\frac{2k}{2\pi} \theta_y \theta_z)} \right) \\
        =&\frac{1}{2\pi}\int_0^{2\pi} d\theta_z\left(\frac{1}{2\pi}\sum_{n''}e^{ -i n''(\theta_x+\theta_y)} \right) e^{-i \theta_z (n+n'-\frac{2k}{2\pi}(\theta_x+\theta_y))}\\
        =&\frac{1}{2\pi}\int_0^{2\pi} d\theta_z \delta_{2\pi}(\theta_x+\theta_y) e^{i \theta_z (n+n'-\frac{2k}{2\pi}(\theta_x+\theta_y))}\\
        =&\delta_{x,y^{-1}} \delta_{n',-n+2k\left(\frac{\theta_x+\theta_y}{2\pi}\right)} = C_{(x,n),(y,n')}\,.
    \end{split}
    \end{equation}
And for $(ST)^3$, one first has
\begin{equation}
    (ST)_{(x,n),(y,n')} = \frac{1}{2\pi}e^{-i(n\theta_y + n' \theta_x -\frac{2k}{2\pi}\theta_x \theta_y)} e^{i(n' \theta_y - \frac{k}{2\pi} \theta_y^2)}\,,
\end{equation}
then one can check
\begin{equation}
    \begin{split}
        & (ST)^3_{(x,n),(y,n')}\\
        =&\frac{1}{(2\pi)^3} \int d \theta_{z_1} d\theta_{z_2} \sum_{n''_1,n''_2} e^{-i(n\theta_{z_1} + n''_1 \theta_x -\frac{2k}{2\pi}\theta_x \theta_{z_1})} e^{-i(n''_{1}\theta_{z_2} + n''_2 \theta_{z_1} -\frac{2k}{2\pi}\theta_{z_1} \theta_{z_2})} e^{-i(n''_{2}\theta_{y} + n' \theta_{z_2} -\frac{2k}{2\pi}\theta_{z_2} \theta_{y})}\\
        &\times e^{i (n''_1 \theta_{z_1} + n''_2 \theta_{z_2}+n' \theta_y)} e^{-i\frac{k}{2\pi}(\theta_{z_1}^2 + \theta_{z_2}^2 + \theta_{y}^2)}\\
        =&\frac{1}{2\pi} \int d\theta_{z_1} d\theta_{z_2} \delta_{2\pi}(\theta_x + \theta_{z_2} - \theta_{z_1}) \delta_{2\pi} (\theta_{z_1}+\theta_y-\theta_{z_2}) e^{-i(n \theta_{z_1} + n' \theta_{z_2}-n' \theta_y)} \\
        &\times e^{-i \frac{k}{2\pi}(\theta_{z_1}^2 + \theta_{z_2}^2 + \theta_y^2-2\theta_x \theta_{z_1}-2\theta_{z_1}\theta_{z_2}-\theta_{z_2}\theta_y)}\\
        =&\frac{1}{2\pi} \int d \theta_{z_1} d\theta_{z_2} \delta_{2\pi}(\theta_x + \theta_{z_2} - \theta_{z_1}) \delta_{2\pi} (\theta_{z_1}+\theta_y-\theta_{z_2}) e^{-i \theta_{z_1}(n+n')} e^{i \frac{2k}{2\pi}\theta_{z_1}(\theta_x + \theta_y)}\\
        =&\frac{1}{2\pi} \int d \theta_{z_1} \delta_{2\pi}(\theta_x + \theta_y) e^{-i \theta_{z_1}(n+n'-\frac{2k}{2\pi} (\theta_x+\theta_y))}\\
        =& \delta_{x,y^{-1}} \delta_{n',-n+2k\left(\frac{\theta_x+\theta_y}{2\pi}\right)}=C_{(x,n),(y,n')}\,.
    \end{split}
\end{equation}


\subsection*{Proof of fusion coefficient}
The fusion coefficient can be obtained by considering the Verlinde formula

    \begin{equation}
        \begin{split}
        &N_{(x,n),(y,n')}^{(z,n'')} \\
        =&\int_{0}^{2\pi} d\theta_{w} \sum_{m} \frac{S_{(x,n),(w,m)} S_{(y,n'),(w,m)} S^*_{(z,n''),(w,m)}}{S_{(1,0),(w,m)}} \\
        =&\frac{1}{(2\pi)^2} \int_{0}^{2\pi} d\theta_w \sum_m e^{-i(n \theta_w+m\theta_x-\frac{2k}{2\pi}\theta_x \theta_w)}e^{-i(n' \theta_w+m\theta_y-\frac{2k}{2\pi}\theta_y \theta_w)}e^{i(n'' \theta_w+m\theta_z-\frac{2k}{2\pi}\theta_z \theta_w)} \\
        =&\frac{1}{2\pi} \int_0^{2\pi} d\theta_w \delta_{2\pi} (\theta_x + \theta_y - \theta_z) e^{-i(n+n'-n'')\theta_w} e^{i \frac{2k}{2\pi} (\theta_x+\theta_y-\theta_z)\theta_w}\\
        =& \delta_{xy,z}\delta_{n'',n+n'- 2k \left(\frac{\theta_x+\theta_y-\theta_z}{2\pi}\right)}
        \end{split}
    \end{equation}

\subsection*{Check of Lagrangian algebras}

We will check the objects $\mathcal{L}_q^k$ 
    \begin{equation}
        \mathcal{L}^k_q = \bigoplus_m \bigoplus_{n=0}^{q-1} W_{\left(e^{\frac{2\pi i}{q}n},\frac{k}{q}n+qm\right)}\,.
    \end{equation}
is a Lagrangian algebra. One can show that the component $W_{\left(e^{\frac{2\pi i}{q}n},\frac{k}{q}n+qm\right)}$ has topological spin zero by the $T$-matrix
    \begin{equation}
        \begin{split}
        &T_{\left(e^{\frac{2\pi i}{q}n},\frac{k}{q}n+qm\right),\left(e^{\frac{2\pi i}{q}n},\frac{k}{q}n+qm\right)}\\ =& e^{i\left((\frac{k}{q}n+qm)(\frac{2\pi}{q}n) -\frac{k}{2\pi}(\frac{2\pi}{q}n)^2\right)} = 1\,,
        \end{split}
    \end{equation}
where we omit the delta function $\delta_{x,y} = \delta_{2\pi}(0)$ when $x=y=e^{\frac{2\pi i}{q}n}$, and we use
    \begin{equation}
        (\frac{k}{q}n+qm)(\frac{2\pi}{q}n) -\frac{k}{2\pi}(\frac{2\pi}{q}n)^2=2\pi mn \in 2\pi \mathbb{Z}\,.
    \end{equation}
Moreover, one can check that they braid trivially via $S$-matrix 
    \begin{equation}
        \begin{split}
        &S_{\left(e^{\frac{2\pi i}{q}n},\frac{k}{q}n+qm\right),\left(e^{\frac{2\pi i}{q}n'},\frac{k}{q}n'+qm'\right)}\\
        =& e^{-i \left((\frac{k}{q}n+qm)(\frac{2\pi}{q}n') + (\frac{k}{q}n'+qm')(\frac{2\pi}{q}n)-\frac{2k}{2\pi} (\frac{2\pi}{q}n)(\frac{2\pi}{q}n') \right)}=1\,,
        \end{split}
    \end{equation}
where we use
    \begin{equation}
        \begin{split}
            &(\frac{k}{q}n+qm)(\frac{2\pi}{q}n') + (\frac{k}{q}n'+qm')(\frac{2\pi}{q}n)-\frac{2k}{2\pi} (\frac{2\pi}{q}n)(\frac{2\pi}{q}n')\\
            =&\frac{4\pi k}{q^2}n n'+2\pi (nm'+n'm) -\frac{4\pi k}{q^2} n n'=2\pi (nm'+n'm)\in 2\pi \mathbb{Z}\,.
        \end{split}
    \end{equation}

One can do similar check for fermionic Lagrangian algebra. When $k=2\ \textrm{mod}\ 4$ one has
    \begin{equation}
        \mathcal{L}_{f} = \bigoplus_{m}\bigoplus_{n=0,1} W_{((-1)^n,2m)}\,,
    \end{equation}
and when $k=0\ \textrm{mod}\ 4$ one has
    \begin{equation}
        \mathcal{L}_{f} = \bigoplus_{m}\bigoplus_{n=0,1} W_{((-1)^n,2m+n)}\,.
    \end{equation}
Then one can check
    \begin{equation}
        T_{((-1)^n,2m),((-1)^n,2m)} = (-1)^{\frac{k}{2}n^2}\,,\quad T_{((-1)^n,2m+n),((-1)^n,2m+n)} = (-1)^{\left(\frac{k}{2}-1\right)n^2}\,,
    \end{equation}
and
    \begin{equation}
        S_{((-1)^n,2m),((-1)^{n'},2m')} = S_{((-1)^n,2m+n),((-1)^{n'},2m'+n')}=1\,.
    \end{equation}


\section{Properties of $S$-matrix for $G=SU(2)$} \label{sec:appE}

First we will check the formula in \eqref{SU(2)-S-properties}.

\subsection*{Calculation of $\frac{1}{4\pi}\int_{0}^{2\pi} d r' \sum_{n'}S_{(r,n),(r',n')}S_{(r',n'),(r'',n'')}$}

\begin{equation}
    \begin{split}
    &\frac{1}{4\pi}\int_{0}^{2\pi} d r' \sum_{n'}S_{(r,n),(r',n')}S_{(r',n'),(r'',n'')}\\ =&  \frac{1}{4\pi}\int_{0}^{2\pi} d r' \sum_{n'} 4\cos \frac{1}{2}(n r' + n' r) \cos\frac{1}{2}(n' r'' + n'' r')\\
    =& \frac{1}{4\pi}\int_{0}^{2\pi} d r' \sum_{n'} \left(e^{\frac{i}{2}(n r' + n' r)} + e^{-\frac{i}{2}(n r' + n' r)} \right) \left( e^{\frac{i}{2}(n' r'' + n'' r')} + e^{-\frac{i}{2}(n' r'' + n'' r')} \right)\\
    =& \frac{1}{4\pi}\int_{0}^{2\pi} d r' \sum_{n'}\left(e^{\frac{i}{2} n' (r + r'')+\frac{i}{2} r' (n+n'')} + e^{-\frac{i}{2} n' (r + r'')-\frac{i}{2} r' (n+n'')} + e^{\frac{i}{2} n' (r - r'')+\frac{i}{2} r' (n-n'')} + e^{-\frac{i}{2} n' (r - r'')-\frac{i}{2} r' (n-n'')}\right)\\
    =& \frac{1}{2} \int_{0}^{2\pi} dr' \left(\delta_{2\pi}(\frac{1}{2}r + \frac{1}{2}r'')(e^{\frac{i}{2}r'(n+n'')}+e^{-\frac{i}{2}r'(n+n'')}) + \delta_{2\pi}(\frac{1}{2}r - \frac{1}{2}r'')(e^{\frac{i}{2}r'(n-n'')}+e^{-\frac{i}{2}r'(n-n'')})  \right)\\
    =&\frac{1}{2} \int_{-2\pi}^{2\pi} dr'\left(\delta_{2\pi}(\frac{1}{2}r + \frac{1}{2}r'')e^{\frac{i}{2}r'(n+n'')} + \delta_{2\pi}(\frac{1}{2}r - \frac{1}{2}r'')e^{\frac{i}{2}r'(n-n'')} \right)\\
    =& \int_{-\pi}^{\pi} d\tilde{r} \delta_{2\pi}(\frac{1}{2}r - \frac{1}{2}r'')e^{i \tilde{r}(n-n'')} = 2\pi \delta_{2\pi}(\frac{1}{2}r - \frac{1}{2}r'') \delta_{n,n''}
    \end{split}
\end{equation}
where $\delta_{2\pi}(\frac{1}{2}r+\frac{1}{2}r'')=0$ since we choose $r,r''\in (0,2\pi)$.

\subsection*{Calculation of $\frac{1}{4\pi}\int_0^{2\pi} dr' \sum_{n'} S_{(0,j),(r',n')} S_{(r',n'),(r,n)}$}
    \begin{equation}
    \begin{split}
        &\frac{1}{4\pi}\int_0^{2\pi} dr' \sum_{n'} S_{(0,j),(r',n')} S_{(r',n'),(r,n)}\\
        =& \frac{1}{2\pi}\int_{0}^{2\pi} dr' \sum_{n'} \frac{\sin \frac{(j+1)}{2} r'}{\sin \frac{1}{2}r'} \cos\frac{1}{2}(n r'+n'r)\\
        =&\frac{1}{4\pi}\int_{0}^{2\pi} dr' \sum_{n'} \frac{\sin \frac{(j+1)}{2} r'}{\sin \frac{1}{2}r'} \left(e^{i\frac{1}{2}(n r'+n'r)}+e^{-i\frac{1}{2}(n r'+n'r)}\right)
    \end{split}
    \end{equation}
The summation over $n'$ will impose $r=0$, but this contradicts with $0<r<2\pi$. Therefore, the expression should vanish.

\subsection*{Calculation of $\int_0^{2\pi}  (\frac{1}{\pi} \sin^2 \frac{r'}{2})d r' \sum_n S_{(0,j),(r,n)} S_{(r,n),(0,j')}$}
    \begin{align}
        &\int_0^{2\pi}  (\frac{1}{\pi} \sin^2 \frac{r'}{2})d r' \sum_n S_{(0,j),(r,n)} S_{(r,n),(0,j')}\nonumber\\
        =&\int_0^{2\pi} d r \sum_n (\frac{1}{\pi} \sin^2 \frac{r}{2}) \frac{\sin \frac{1}{2}(j+1) r}{\sin \frac{1}{2}r} \frac{\sin\frac{1}{2}(j'+1)r}{\sin \frac{1}{2}r}\nonumber\\
        =&\sum_n \frac{1}{\pi} \int_0^{2\pi} dr \sin \left(\frac{j+1}{2} r\right) \sin \left(\frac{j'+1}{2}r\right)\nonumber\\
        =&\sum_n \frac{1}{2\pi} (\int_0^{2\pi} dr \cos\left(\frac{j-j'}{2}r\right) -\int_0^{2\pi} dr \cos(\frac{j+j'+2}{2}r)\,.
    \end{align}
The second term vanishes
    \begin{equation}
        \int_0^{2\pi} dr \cos(\frac{j+j'+2}{2}r)=\frac{2}{j+j'+2} \sin \left(\frac{j+j'+2}{2} r\right) \Big|_{0}^{2\pi}=0\,,
    \end{equation}
since $j+j'+2\neq0$. Therefore, we only have the first term, which is only non-zero when $j=j'$. The result is
    \begin{equation}
        \begin{split}
        &\int_0^{2\pi}  (\frac{1}{\pi} \sin^2 \frac{r'}{2})d r' \sum_m S_{(0,j),(r,n)} S_{(r,n),(0,j')}\\
        =& \left(\sum_n 1\right) \delta_{j,j'} = \left(\sum_{n} e^{0 i n } \right) \delta_{j,j'} = 2\pi\delta_{2\pi}(0) \delta_{j,j'}
        \end{split}
    \end{equation}

\subsection*{Calculation of $\int_0^{2\pi}  (\frac{1}{\pi} \sin^2 \frac{r'}{2})d r' \sum_m S_{(2\pi,j),(r,n)} S_{(r,n),(0,j')}$}
The calculation is basically the same to the previous one and we get
    \begin{equation}
        \begin{split}
            &\int_0^{2\pi}  (\frac{1}{\pi} \sin^2 \frac{r'}{2})d r' \sum_m S_{(2\pi,j),(r,n)} S_{(r,n),(0,j')}\\
            =&\sum_n (-1)^n \delta_{j,j'}= 2\pi \delta_{2\pi}(\pi) \delta_{j,j'} =0\,.
        \end{split}
    \end{equation}

\subsection*{Fusion rule}
The fusion coefficient considered in \eqref{SU(2)-fusion-generic} is
    \begin{equation}
        \begin{split}
        &N_{(r_1,n_1),(r_2,n_2)}^{(r_3,n_3)} =\frac{1}{4\pi} \int_{0}^{2\pi} dr \sum_{n} \frac{S_{(r_1,n_1),(r,n)}S_{(r_2,n_2),(r,n)}S^*_{(r_3,n_3),(r,n)}}{S_{(0,0),(r,n)}}\\
        =& \frac{1}{4\pi}\int_{0}^{2\pi} dr \sum_n \left(e^{\frac{i}{2}(n_1r+r_1 n)}+e^{-\frac{i}{2}(n_1r+r_1 n)}\right)\left(e^{\frac{i}{2}(n_2r+r_2 n)}+e^{-\frac{i}{2}(n_2r+r_2 n)}\right)\left(e^{\frac{i}{2}(n_3r+r_3 n)}+e^{-\frac{i}{2}(n_3r+r_3 n)}\right)\\
        =& \frac{1}{2} \int_{0}^{2\pi} dr \delta_{2\pi}(\frac{1}{2}r_1+\frac{1}{2}r_2+\frac{1}{2}r_3) (e^{\frac{i}{2}r(n_1+n_2+n_3)}+e^{-\frac{i}{2}r(n_1+n_2+n_3)})\\
        +&\frac{1}{2} \int_{0}^{2\pi} dr \delta_{2\pi}(\frac{1}{2}r_1+\frac{1}{2}r_2-\frac{1}{2}r_3) (e^{\frac{i}{2}r(n_1+n_2-n_3)}+e^{-\frac{i}{2}r(n_1+n_2-n_3)})\\
        +&\frac{1}{2} \int_{0}^{2\pi} dr \delta_{2\pi}(\frac{1}{2}r_1-\frac{1}{2}r_2+\frac{1}{2}r_3) (e^{\frac{i}{2}r(n_1-n_2+n_3)}+e^{-\frac{i}{2}r(n_1-n_2+n_3)})\\
        +&\frac{1}{2} \int_{0}^{2\pi} dr \delta_{2\pi}(-\frac{1}{2}r_1+\frac{1}{2}r_2+\frac{1}{2}r_3) (e^{\frac{i}{2}r(-n_1+n_2+n_3)}+e^{-\frac{i}{2}r(-n_1+n_2+n_3)})
        \end{split}
    \end{equation}
Since $r_1,r_2,r_3 \in (0,2\pi)$ and we further assumed $r_1 >r_2$, then one has
    \begin{equation}
        \delta_{2\pi}(\frac{1}{2}r_1+\frac{1}{2}r_2+\frac{1}{2}r_3)=\delta_{2\pi}(-\frac{1}{2}r_1+\frac{1}{2}r_2+\frac{1}{2}r_3)=0\,,
    \end{equation}
and we have
    \begin{equation}
    \begin{split}
        &\frac{1}{4\pi}\int_{0}^{2\pi} dr \sum_{n} \frac{S_{(r_1,n_1),(r,n)}S_{(r_2,n_2),(r,n)}S^*_{(r_3,n_3),(r,n)}}{S_{(0,0),(r,n)}}\\
        =&\frac{1}{2} \int_{0}^{2\pi} dr \delta_{2\pi}(\frac{1}{2}r_1+\frac{1}{2}r_2-\frac{1}{2}r_3) (e^{\frac{i}{2}r(n_1+n_2-n_3)}+e^{-\frac{i}{2}r(n_1+n_2-n_3)})\\
        +&\frac{1}{2} \int_{0}^{2\pi} dr \delta_{2\pi}(-\frac{1}{2}r_1+\frac{1}{2}r_2+\frac{1}{2}r_3) (e^{\frac{i}{2}r(-n_1+n_2+n_3)}+e^{-\frac{i}{2}r(-n_1+n_2+n_3)})\\
        =&\frac{1}{2} \int_{-2\pi}^{2\pi} dr \delta_{2\pi}(\frac{1}{2}r_1+\frac{1}{2}r_2-\frac{1}{2}r_3) e^{\frac{i}{2}r(n_1+n_2-n_3)} + \frac{1}{2} \int_{-2\pi}^{2\pi} dr \delta_{2\pi}(-\frac{1}{2}r_1+\frac{1}{2}r_2+\frac{1}{2}r_3) e^{\frac{i}{2}r(-n_1+n_2+n_3)}\\
        =& \int_{-\pi}^{\pi} d\tilde{r} \delta_{2\pi}(\frac{1}{2}r_1+\frac{1}{2}r_2-\frac{1}{2}r_3) e^{i\tilde{r}(n_1+n_2-n_3)} + \int_{-\pi}^{\pi} d\tilde{r} \delta_{2\pi}(-\frac{1}{2}r_1+\frac{1}{2}r_2+\frac{1}{2}r_3) e^{i\tilde{r}(-n_1+n_2+n_3)}\\
        =&2\pi \delta_{2\pi}(\frac{1}{2}r_1+\frac{1}{2}r_2-\frac{1}{2}r_3) \delta_{n_3,n_1+n_2} + 2\pi \delta_{2\pi}(-\frac{1}{2}r_1+\frac{1}{2}r_2+\frac{1}{2}r_3) \delta_{n_3,n_1-n_2}\,.
    \end{split}
    \end{equation}

\bibliographystyle{JHEP}
\bibliography{refs}

\end{document}